\documentclass[%12pt,
onecolumn, % triplespace,
journal]{IEEEtran}

\usepackage{rotating}
\usepackage{amsmath}
\usepackage{graphicx,amssymb}
\usepackage{algorithm}
\usepackage{algorithmic}
\usepackage{amsthm,color}
\usepackage{diagbox}
\usepackage{multirow}
\usepackage{subcaption}
\usepackage{makecell}
\usepackage{hyperref}
\usepackage{soul}
\usepackage{mdframed}
\usepackage{algorithm}
\usepackage{makecell}

\usepackage[affil-it]{authblk}

\makeatletter

\@addtoreset{algorithm}{section}
\makeatother

\newtheorem{thm}{Theorem}[section]

\newtheorem{remark}[thm]{Remark}

\renewcommand{\theequation} {\thesection.\arabic{equation}}

\numberwithin{equation}{section}

\renewcommand{\thesection}{\arabic{section}}
\setstcolor{red}
\begin{document}

\title
{Wiener   filters on graphs and distributed  polynomial approximation algorithms}

\author{Cong Zheng, Cheng Cheng, and Qiyu Sun
\thanks{Zheng and Sun are with the  Department of Mathematics,
University of Central Florida,
Orlando, Florida 32816;   Cheng is
with the School of Mathematics, Sun Yat-sen University, Guangzhou,
Guangdong 510275, China.
Emails:  acongz@knights.ucf.edu; chengch66@mail.sysu.edu.cn;  qiyu.sun@ucf.edu.
This work is partially supported by the National
Science Foundation (DMS-1816313), National Nature Science Foundation
of China (12171490) and Guangdong Province Nature Science
Foundation (2022A1515011060)
}
}

\maketitle

\begin{abstract}
In this paper, we consider  Wiener filters to reconstruct deterministic and (wide-band) stationary graph signals from their  observations corrupted by random noises,
and we propose  distributed algorithms to implement
 Wiener filters and
inverse filters %associated with a polynomial filter of  commutative graph shifts
on networks in which agents are equipped with a data processing
subsystem  for limited data storage and computation power,
and with  a  one-hop communication subsystem   for  direct data exchange only with their adjacent agents.
The proposed distributed polynomial approximation algorithm is an  exponential convergent quasi-Newton method based on  Jacobi polynomial approximation and  Chebyshev interpolation polynomial approximation to analytic functions on a cube.
% and it becomes the  gradient descent algorithm when the polynomial approximation filter is appropriately chosen.
Our numerical simulations
show that Wiener filtering procedure performs
better on denoising (wide-band)  stationary signals than the Tikhonov regularization approach does, and that the proposed polynomial approximation algorithms
converge faster
than the Chebyshev polynomial approximation algorithm
 and gradient decent algorithm  % with optimal step size
 do in the implementation of an inverse filtering procedure
associated with a polynomial filter of  commutative graph shifts.
   \end{abstract}


%\vspace{-1em}

\section{Introduction}
 Massive data sets on networks are collected in numerous applications, such as  (wireless) sensor networks, smart
grids and social networks  \cite{Wass94}-\cite{Cheng19}.
%\cite{Wass94, chong2003,  mfa07, Yick08, Motee17, Hebner17, Cheng19}.
Graph signal processing provides an innovative framework
to extract knowledge from (noisy) data sets residing on networks %and many other irregular domains
\cite{sandryhaila13}-\cite{ncjs22}.
%\cite{sandryhaila13, shuman13, sandryhaila14,  bronstein17,  Ortega18, stankovic2019introduction, dong20, ncjs22}.
 Graphs ${\mathcal G} = (V, E)$  are widely used to model the complicated   topological structure of networks in engineering applications, where
a vertex in $V$ may represent an agent of the network and an edge in $E$ between vertices could indicate that the
corresponding agents  have a
 peer-to-peer communication link between
them
and/or they are within certain range in the spatial space. In this paper, we consider distributed implementation of
 Wiener filtering procedure and inverse filtering procedure on  simple graphs   (i.e.,
 unweighted undirected graphs containing no  loops or multiple edges)
   of large order $N\ge 1$.

 Many data sets on a network can be considered as signals  ${\bf x}=(x_i)_{i\in V}$ residing on the graph ${\mathcal G}$, % ${\mathcal G}=(V, E)$,
 where $x_i$ represents the real/complex/vector-valued data at the vertex/agent $i\in V$.
In this paper, the data $x_i$ at each vertex $i\in V$ is assumed to be real-valued.  The filtering procedure for signals on a network is a linear transformation
    \begin{equation}\label{filtering.def}
{\bf x}\longmapsto {\bf y}={\bf H}{\bf x}, \end{equation}
which maps a graph signal   ${\bf x}$
 to another graph signal ${\bf y}={\bf H}{\bf x}$, and ${\bf H}=(H(i,j))_{ i,j\in V}$ is known as a {\em graph filter}.
 In this paper, we assume that graph filters are real-valued.

We say that a matrix ${\bf S}=(S(i,j))_{i,j\in V}$  on  the  graph ${\mathcal G}=(V, E)$ is a {\em graph shift} if
$S(i,j)\ne 0$ only if either $j=i$ or $(i,j)\in E$.
 Graph shift  is a basic concept in graph signal processing, and illustrative examples are the adjacency matrix ${\bf A}$, Laplacian matrix ${\bf L}={\bf D}-{\bf A}$, and
 symmetrically normalized  Laplacian ${\bf L}^{\rm sym}:={\bf D}^{-1/2} {\bf L}{\bf D}^{-1/2}$, where ${\bf D}$
 is the degree matrix of the graph
  \cite{sandryhaila13}, \cite{ncjs22}-\cite{jiang19}.
% \cite{sandryhaila13,   ncjs22,  segarra17, Coutino17, jiang19}.
In \cite{ncjs22},   the notion of multiple commutative graph shifts ${\bf S}_1, \dots,  {\bf S}_d$ are introduced,
%	\vspace{-0.4em}
\begin{equation}\label{commutativityS}
	{\bf S}_k{\bf S}_{k'}={\bf S}_{k'}{\bf S}_k,\  1\le k,k'\le d,
%	\vspace{-0.6em}
\end{equation}
and some multiple commutative graph shifts on circulant/Cayley graphs and on Cartesian product graphs  are constructed with physical interpretation.
% One may verify that graph shifts ${\bf S}_1, \ldots, {\bf S}_d$ are commutative  if they can be diagonalized simultaneously, i.e.,
%there exists a nonsingular matrix ${\bf P}$ such that ${\bf P}^{-1} {\bf S}_k{\bf P}, 1\le k\le d$, are diagonal matrices,
%and the converse holds if ${\bf S}_1, \ldots, {\bf S}_d$ are symmetric.
%Commutative graph shifts are usually designed/selected to have specific features
%and physical interpretation.
An important property for   commutative graph shifts ${\mathbf S}_1, \ldots, {\mathbf S}_d$ is that they
can be upper-triangularized  simultaneously,
\begin{equation}
	\label{upperdiagonalization}
\widehat{\bf S}_k={\bf U}^{\rm H}{\bf S}_k{\bf U},\  1\le k\le d,
	\end{equation}
where ${\bf U}$ is a unitary matrix, ${\bf U}^{\rm H}$ is the Hermitian of the matrix ${\bf U}$,  and
	$\widehat{\bf S}_k=(\widehat S_{k}(i,j))_{1\le i, j\le N}, 1\le k\le d$, are upper triangular matrices
 \cite[Theorem 2.3.3]{horn1990matrix}.
As
 $\widehat{S}_k(i, i), 1\le i\le N$, are eigenvalues of ${\bf S}_k, 1\le k\le d$,
 we call the set	
 \begin{equation}\label{jointspectrum.def} \Lambda=\big\{\pmb \lambda_i=\big(\widehat{S}_1(i,i), ..., \widehat{ S}_d(i,i)\big), 1\le i\le N\big\}\end{equation}
as the {\em joint spectrum} of  ${\bf S}_1, \ldots, {\bf S}_d$  \cite{ncjs22}.
  For the case that graph shifts ${\bf S}_1, \ldots, {\bf S}_d$ are symmetric, one may verify that their joint spectrum are contained in some cube,
%$[{\pmb \mu}, {\pmb \nu}]=[\mu_1,\nu_1] \times \cdots \times[\mu_d,\nu_d]$   in ${\mathbb R}^d$, i.e.,
\begin{equation}\label{jointspectralcubic.def}
\Lambda\subset [{\pmb \mu}, {\pmb \nu}]:=[\mu_1,\nu_1] \times \cdots \times[\mu_d,\nu_d]\subset {\mathbb R}^d.
\end{equation}

    A popular family of graph filters contains
    {\em polynomial graph filters}  of  commutative graph shifts   ${\bf S}_1, \dots,  {\bf S}_d$,
%	\vspace{-0.6em}
\begin{equation}\label{MultiShiftPolynomial}
	{\bf H}=h({\bf S}_1, \ldots, {\bf S}_d)=\sum_{ l_1=0}^{L_1} \cdots \sum_{ l_d=0}^{L_d}  h_{l_1,\dots,l_d}{\bf S}_1^{l_1}\cdots {\bf S}_d^{l_d},
%	\vspace{-0.6em}
\end{equation}
 %  as its blocks,
   where
$h$ is a multivariate polynomial
in variables $t_1,\cdots,t_d$,
%\vspace{-0.6em}
% \begin{equation*}\label{MultiShiftPolynomial.polynomial}
$$h(t_1, \ldots, t_d)=\sum_{ l_1=0}^{L_1} \cdots \sum_{ l_d=0}^{L_d}  h_{l_1,\dots,l_d} t_1^{l_1} \ldots t_d^{l_d}$$
 % of degree $\sum_{k=1}^d L_k$.
%h_{l_1,\dots,l_d},  0\le l_k\le L_k, 1\le k\le d$ are the polynomial coefficients for the terms  ${\bf S}_1^{l_1}\cdots {\bf S}_d^{l_d}$.
  \cite{ncjs22, segarra17},  \cite{Leus17}-\cite{David2019}.
    %\cite{ncjs22, segarra17,  Leus17, Waheed18, Lu18, shuman18,    mario19,   Emirov19, David2019}.
%By the commutative property \eqref{commutativityS}
%among graph shifts ${\bf S}_1, \dots,  {\bf S}_d$,  one may verify that
%  the polynomial graph filter ${\bf H}$ in \eqref{MultiShiftPolynomial} is
%independent on
%equivalent expressions of the multivariate polynomial $h$ and hence the polynomial filter
%\eqref{MultiShiftPolynomial} is well-defined.
 Commutative graph shifts   ${\bf S}_1, \dots,  {\bf S}_d$ are building blocks for polynomial graph filters and they  play  similar roles  in graph signal processing
   as the  one-order delay $z_1^{-1}, \ldots, z_d^{-1}$ in  multi-dimensional  digital signal processing  \cite{ncjs22}.
  For  polynomial  graph filters  in \eqref{MultiShiftPolynomial},  a significant advantage is that
the corresponding filtering procedure \eqref{filtering.def}
   can be implemented at the vertex level in which  each vertex  is equipped
%with a data processing subsystem for limited data storage and computation power,
%and
with a {\bf one-hop  communication subsystem}, i.e.,  each agent has direct data exchange only with its adjacent  agents, see
   \cite[Algorithms 1 and 2]{ncjs22}.

     Inverse filtering procedure associated with  a polynomial filter
has been widely used in  denoising,  non-subsampled filter banks
 and signal reconstruction, graph semi-supervised
learning and many other applications
\cite{jiang19, Leus17, Lu18}-\cite{Emirov19}, \cite{siheng_inpaint15}-\cite{cheng2021}.
%\cite{jiang19, Leus17, Lu18, shuman18,  mario19,  Emirov19, siheng_inpaint15,    Shi15, sihengTV15, Onuki16, cheng2021}.
%\cite{shuman18, jiang19, mario19, Leus17, Lu18, Emirov19, Shi15, sihengTV15, Onuki16, siheng_inpaint15}.
%(the graph signal sampling and reconstruction is not viewed as the  inverse filtering in general.),
In Sections  \ref{stochasticwienerfilter.section} and \ref{worst-casewienerfilter.section}, we consider   the scenario that
the filtering procedure  \eqref{filtering.def} is associated with a polynomial filter,
 its inputs  ${\bf x}$ %=(x(i))_{i\in V}$ are
 are either
 (wide-band) stationary signals or  deterministic signals with finite energy,
 and
  its outputs  ${\bf y}$ are corrupted by some random noises  which  have mean zero and
their covariance matrix   being   a polynomial filter of graph shifts
% ${\bf S}_1, \ldots, {\mathbf S}_d$, i.e.,
%$ {\mathbb E}{\pmb \epsilon}={\bf 0}$ and ${\bf G}=g({\bf S}_1, \ldots, {\bf S}_d)
%$
%for some multivariate polynomial $g$
\cite{bi2009}-\cite{yagan2020}.
%\cite{bi2009, girault2015, perraudin17, segarrat2017, yagan2020}.
We show that the corresponding stochastic/worst-case Wiener filters   %  ${\bf W}_{\rm wmse}$
are essentially  the product of a polynomial filter and   inverse of another polynomial filter, see  Theorems  \ref{wienerfiltermse.thm}, \ref{widebandwienerfilter.thm} and \ref{wienerfilterworsecase.thm}.  %, and Remark \ref{stochasticwiener.remark}.
  Numerical demonstrations in Sections \ref{randomsignal.demo}
and \ref{denoisingwideband.demo} indicate that the Wiener filtering procedure has better performance on denoising (wide-band) stationary signals  than the conventional Tikhonov
regularization approach does   \cite{ncjs22, Shi15}.

Given a polynomial filter ${\bf H}$ of graph shifts, one
 of the main challenges in the corresponding inverse filtering procedure
\begin{equation}\label{inverseprocedure}
{\bf y}\longmapsto {\bf x}={\bf H}^{-1}{\bf y}
\end{equation}
 is on its distributed implementation, as
  the inverse filter  ${\bf H}^{-1}$ is usually not a polynomial filter of small degree even if ${\bf H}$ is.
%  The inverse filtering procedure can be considered as solving the linear system
%  ${\bf H}{\bf x}={\bf y}$.
The last two authors of this paper proposed the following exponentially convergent quasi-Newton method
 \begin{equation}\label{Approximationalgorithm}
 {\bf e}^{(m)}= {\bf H}{\bf x}^{(m-1)}-{\bf y}\ {\rm and} \
{\bf x}^{(m)}={\bf x}^{(m-1)}-{\bf G}{\bf e}^{(m)}, \ m\ge 1,
\end{equation}
with arbitrary initial ${\bf x}^{(0)}$   to fulfill the inverse filtering procedure,
% where
%%the approximation filter ${\bf G}$ to ${\bf H}^{-1}$  is  of a polynomial form,
%the approximation filter ${\bf G}$ is so chosen that  the spectral radius of ${\bf I}-{\bf G}{\bf H}$  is strictly less than $1$
% \cite{ncjs22, Emirov19, cheng2021}.
 where the polynomial approximation filter ${\bf G}$  to the inverse ${\bf H}^{-1}$
  is  so chosen that
  the spectral radius of ${\bf I}-{\bf G}{\bf H}$  is strictly less than $1$
 \cite{ncjs22, Emirov19, cheng2021}.
    More importantly,  each iteration in  \eqref{Approximationalgorithm}
 includes mainly two  filtering procedures associated with  polynomial filters ${\bf H}$ and ${\bf G}$.  In this paper,
the quasi-Newton method
\eqref{Approximationalgorithm} is used  to implement
the Wiener filtering procedure  and inverse filtering procedure associated with a polynomial filter   on networks whose  agents are equipped
with  %a data processing subsystem for limited data storage and computation power,
 a  one-hop communication subsystem, % for direct data exchange to its adjacent agents only,
 see \eqref{jacobiapproximation.def} and
Algorithms   \ref{Wiener2.algorithm} and \ref{Wiener1.algorithm}.

An important  problem not discussed yet is how to select the polynomial  approximation filter ${\bf G}$ appropriately for  the fast convergence of the
 quasi-Newton method \eqref{Approximationalgorithm}.
The above  problem  has been well studied when
${\bf H}$ is  a polynomial filter  of the  % (symmetrically normalized)
 graph Laplacian (and a single graph shift in general)
 \cite{%sihengTV15,
 Leus17, Emirov19, Shi15, sihengTV15, Shuman18, isufi19}.
% , even though the original polynomial filter ${\bf H}$ and the polynomial approximation filter ${\bf G}$
% may not satisfy the geodesic-width requirement \eqref{width.req},
For a polynomial filter ${\bf H}$
 of multiple  graph shifts,
 optimal/Chebyshev polynomial  approximation filters
are introduced in \cite{ncjs22}.
The construction of Chebyshev polynomial  approximation filters is based
on the exponential approximation property of Chebyshev polynomials to  the reciprocal of a multivariate %{\color{red} non-vanishing or analytic}
polynomial
on the cube containing the joint spectrum of  multiple  graph shifts.
Chebyshev polynomials form a special family of Jacobi polynomials.  In Section \ref{Jacobiapproximation.section},
based on the exponential approximation property of Jacobi polynomials and Chebyshev interpolation polynomials to analytic functions on a cube,
 we introduce
Jacobi polynomial  filters  %  ${\bf G}_M^{(\alpha, \beta)}, M\ge 0$
and Chebyshev interpolation polynomial filters  % ${\bf I}_M, M\ge 0$
  to approximate the inverse filter ${\bf H}^{-1}$, and we use the corresponding quasi-Newton method algorithm  \eqref{jacobiapproximation.def}
  to implement the inverse filtering procedure \eqref{inverseprocedure}.
  Numerical experiments in Section \ref{circulantgraph.demo} indicate that the proposed Jacobi polynomial  approach  with appropriate selection of parameters
  and  Chebyshev interpolation polynomial approach have better performance  than  Chebyshev  polynomial approach  and
   gradient descent method with optimal step size
  do   \cite{ncjs22,jiang19, Leus17, Waheed18, Shi15, sihengTV15, Shuman18, isufi19}.

Notation: Let ${\mathbb Z}_+$ be the set of all nonnegative integers and set
$\mathbb{Z}_+^d=\{(n_1, \ldots, n_d), \ n_k\in {\mathbb Z}_+, 1\le k\le d\}$.
Define    $\|{\bf x}\|_2=(\sum_{i\in V} |x_i|^2)^{1/2}$ for a graph signal ${\bf x}=(x_i)_{i\in V}$
and  $\|{\bf A}\|=\sup_{\|{\bf x}\|_2=1} \|{\bf A}{\bf x}\|_2$ for a graph filter ${\bf A}$.
Denote the transpose of a matrix ${\bf A}$ by
${\bf A}^T$ and  the trace of a square matrix ${\bf A}$ by ${\rm tr}({\bf A})$.
%the maximal/minimum eigenvalue of a symmetric matrix ${\bf A}$ by  $\lambda_{\max}({\bf A})$ and $\lambda_{\min}({\bf A})$ respectively.
% ${\bf A}\preceq {\bf B}$ .
As usual, we use ${\bf O}, {\bf I}, {\bf 0}, {\bf 1}$  to denote the zero matrix, identity matrix,  zero vector and  vector of all $1$s of appropriate sizes respectively.

\section{Preliminaries on Jacobi polynomials and Chebyshev interpolating polynomials}

\begin{figure}[t]
\begin{center}
\includegraphics[width=43mm, height=30mm]
{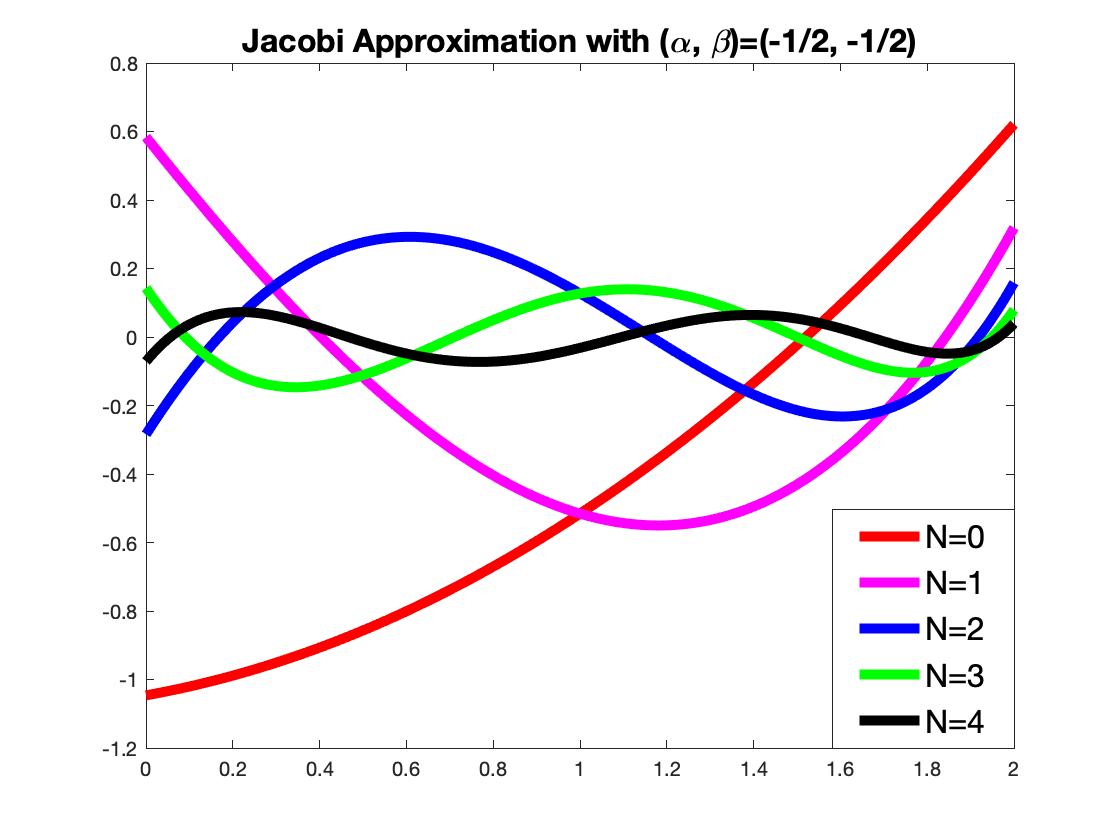}
\includegraphics [width=43mm, height=30mm]{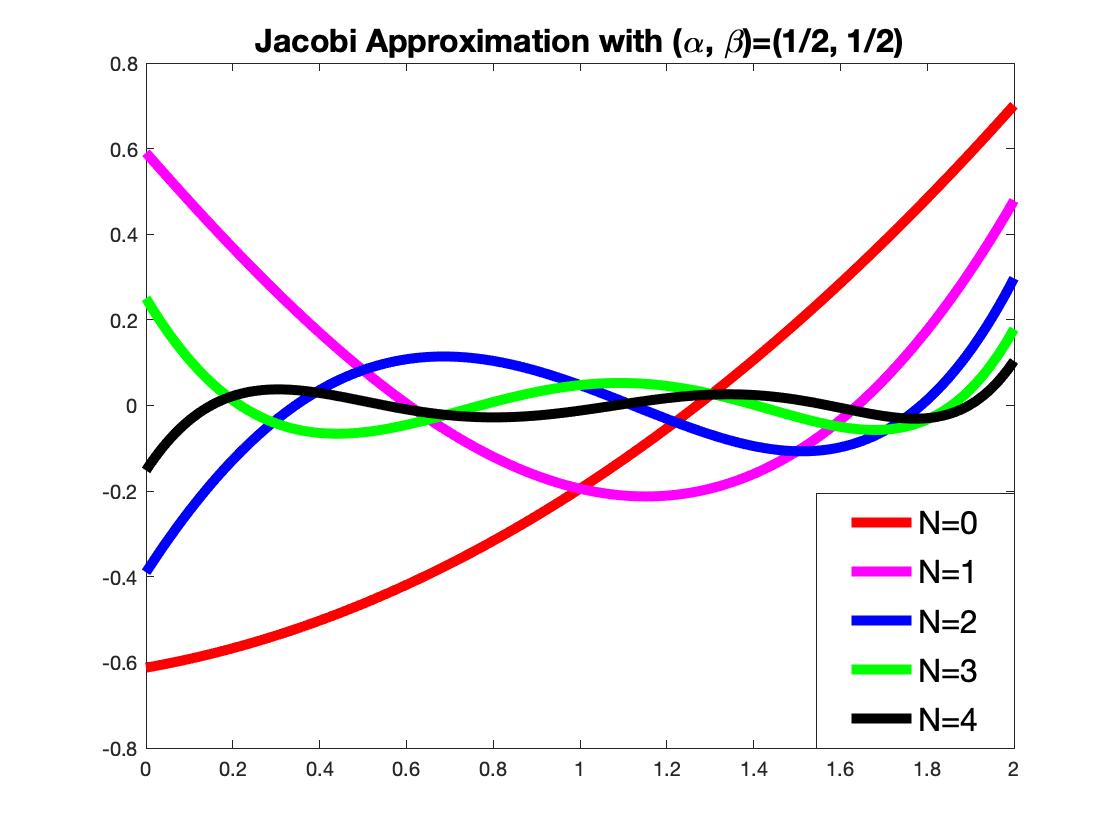}\\
\includegraphics[width=43mm, height=30mm]
{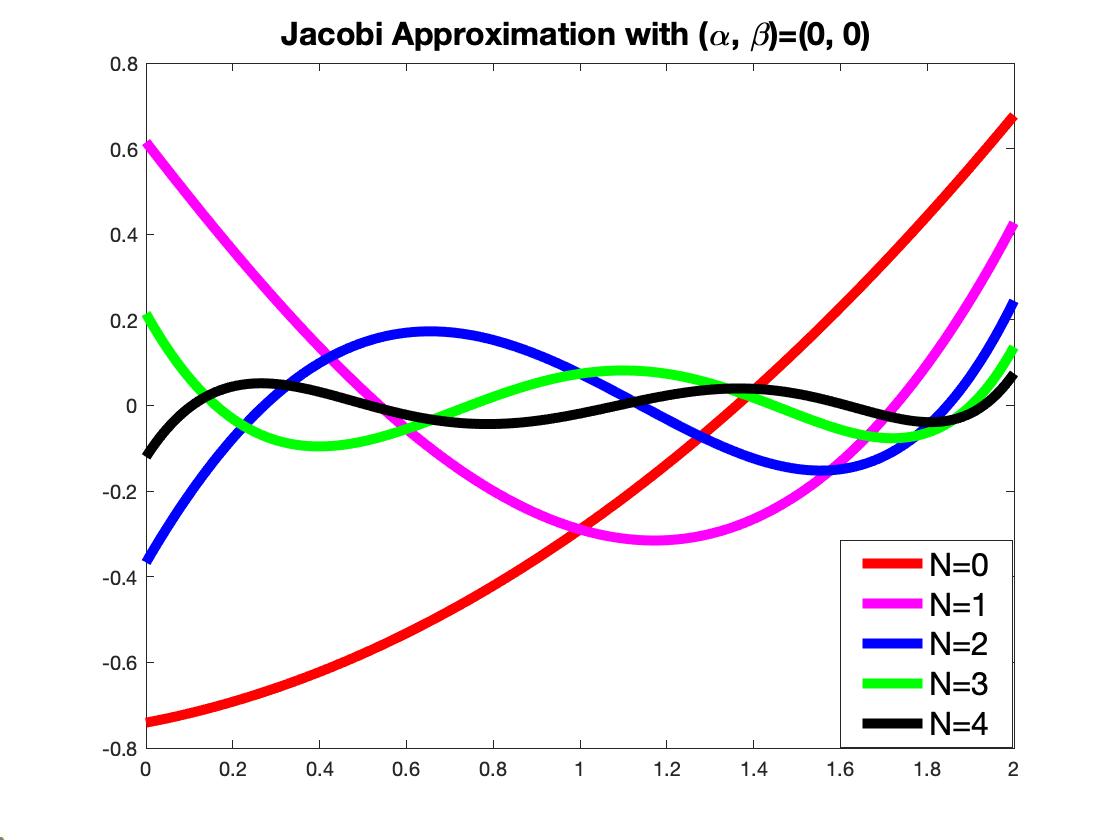}
\includegraphics [width=43mm, height=30mm]{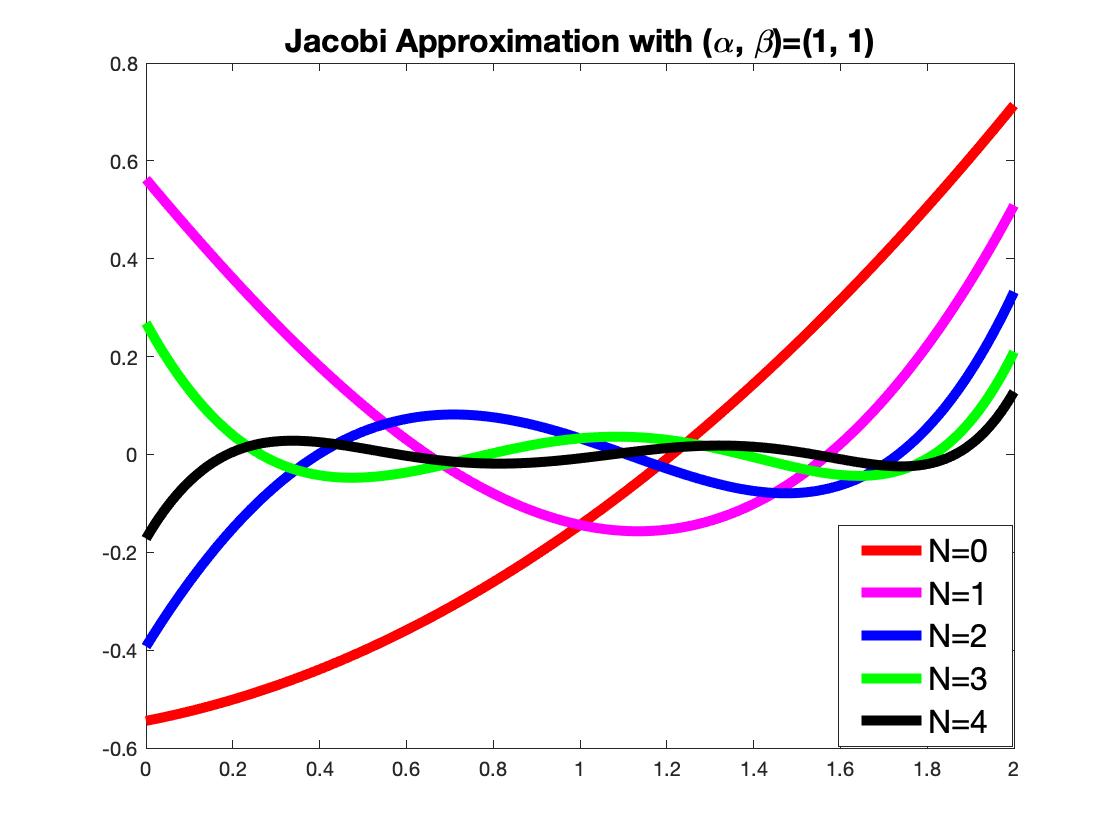}
\\
\includegraphics[width=43mm, height=30mm]
{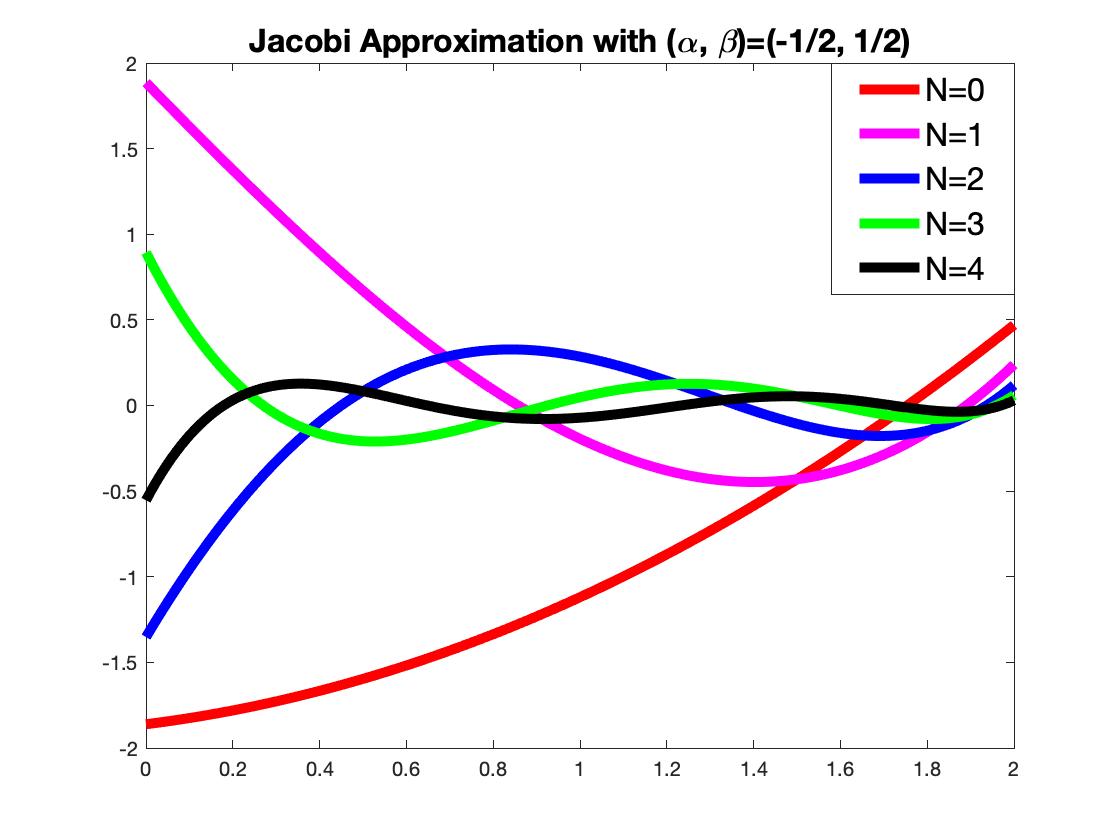}
\includegraphics [width=43mm, height=30mm]{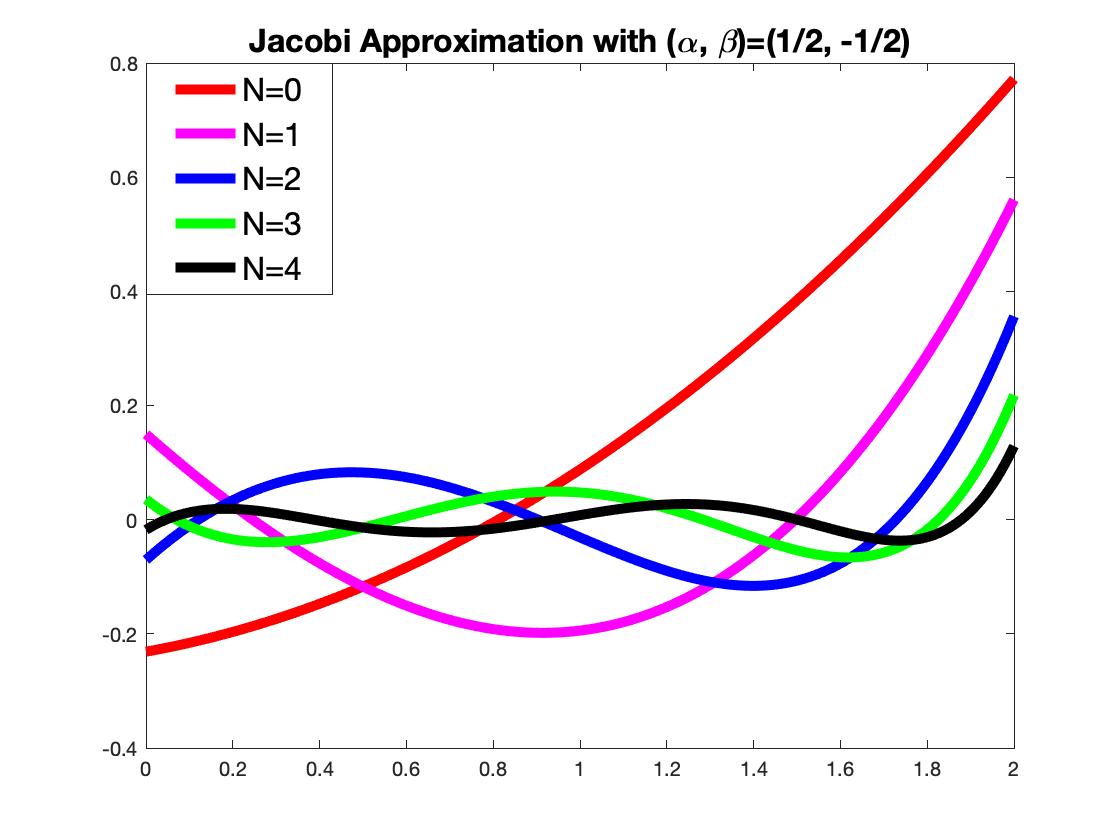}\\
\includegraphics [width=43mm, height=30mm]{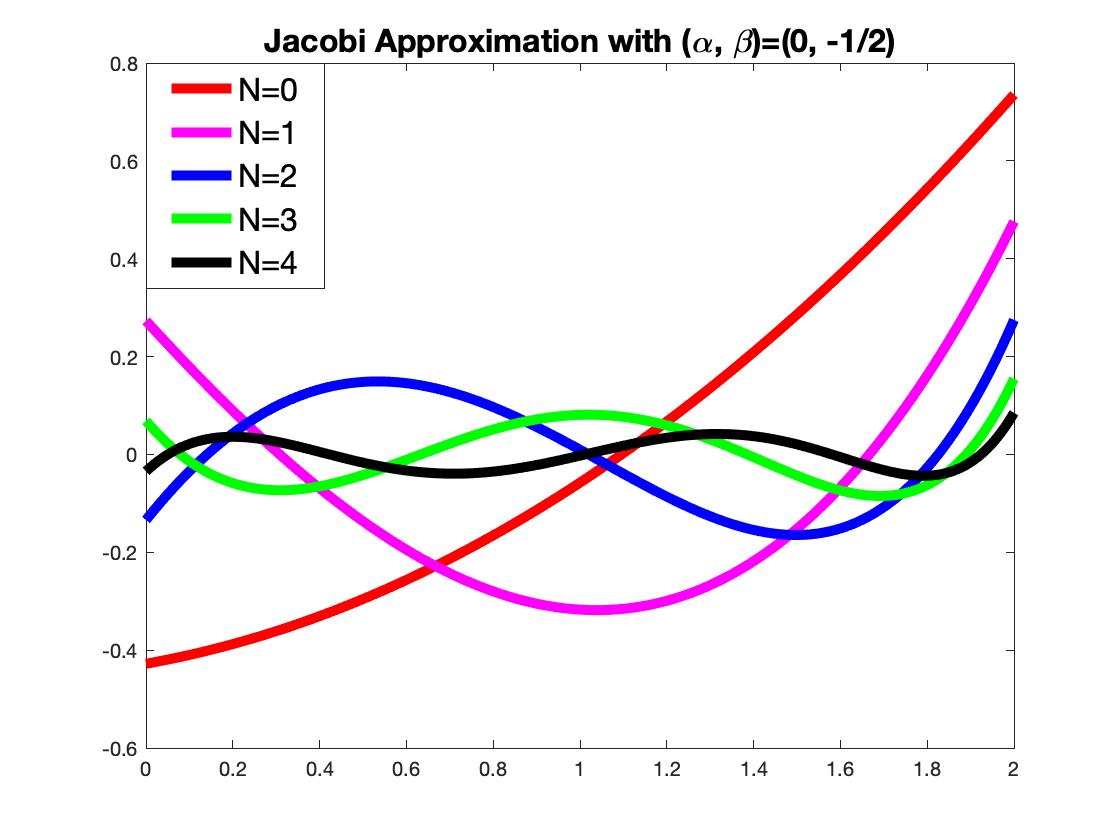}
%\subfigure{\includegraphics [width=38mm, height=30mm]{pic/ZeroNeigth.jpg}}
\includegraphics [width=43mm, height=30mm]{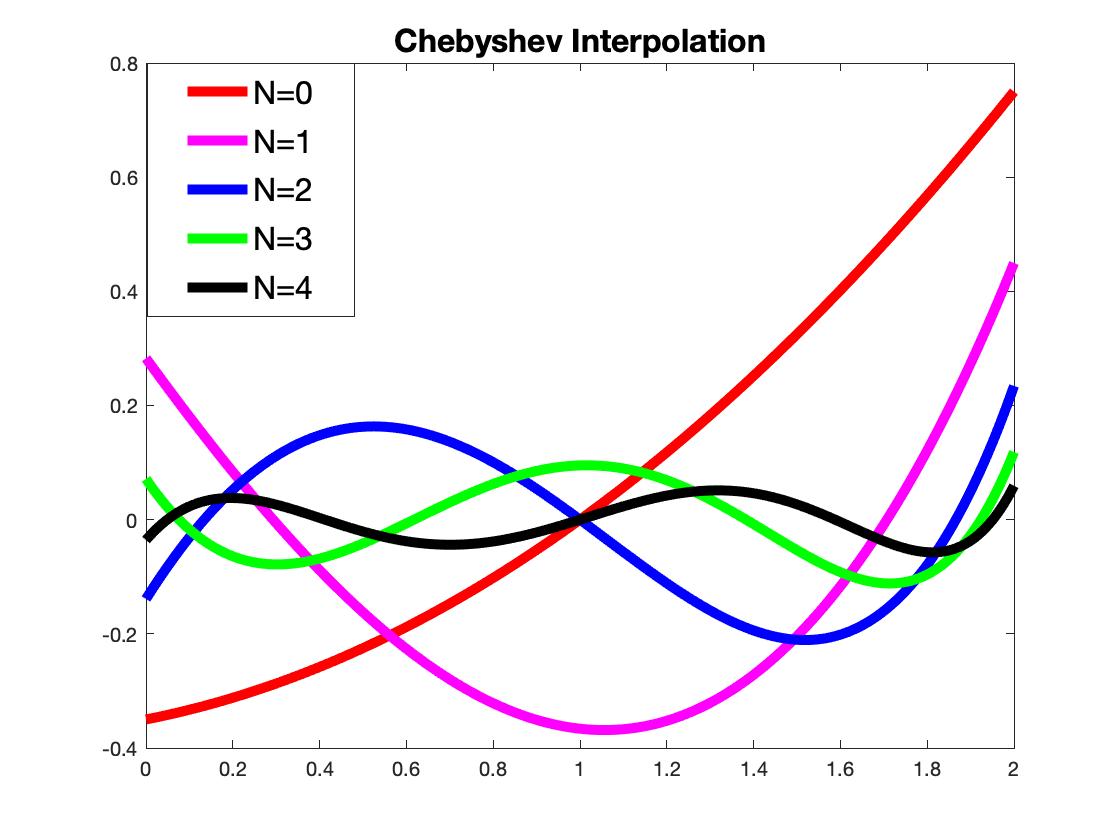}
\caption{Plotted on the top three rows and the left of bottom row
 are the approximation error functions
$1-h_1(t)g_M^{(\alpha, \beta)}(t), t\in [0, 2], 0\le M\le 4$
for  pairs  $(\alpha, \beta)=(-1/2, -1/2)$ (top row left), $(1/2, 1/2)$ (top row right), $(0, 0)$ (second row left),
$(1, 1)$ (second row right),
 $(-1/2, 1/2)$ (third row  left), $(1/2, -1/2)$ (third row  right) and $(0, -1/2)$
(bottom row left). On the bottom row right is the approximation error function
$1-h_1(t)C_M(t), t\in [0, 2], 0\le M\le 4$,
 between the
Chebyshev interpolation polynomial $C_M(t)$ and the reciprocal of the polynomial $h_1(t)$.}
\label{approximation.fig}
\end{center}
\end{figure}

Let  $\alpha, \beta>-1$,
$[{\pmb \mu}, {\pmb \nu}]=[\mu_1,\nu_1] \times \cdots \times[\mu_d,\nu_d]$  be a cube in ${\mathbb R}^d$ with its volume denoted by
$|[{\pmb \mu}, {\pmb \nu}]| $, and  let $h$ be a multivariate polynomial   satisfying
 \begin{equation}\label{polynomial.assump}
    h({\bf t})\neq 0\  \text{ for all }\  {\bf t}\in [{\pmb \mu}, {\pmb \nu}].
\end{equation}
In this section,  we  recall the definitions of
 multivariate Jacobi polynomials  % $P_{{\bf n}; {\pmb \mu}, {\pmb \nu}}^{(\alpha, \beta)}, {\bf n}\in\mathbb{Z}_+^d$
  and
 interpolation  polynomials at Chebyshev nodes,
   and  their exponential approximation property to the reciprocal of the polynomial $h$
 on the cube
$[{\pmb \mu}, {\pmb \nu}]$
 \cite{Ismail2009}-\cite{xiang2012}.
 % \cite{Ismail2009, Shen2011, trefethen2013, wang2011, xiang2012}.
Our numerical simulations indicate that   Jacobi polynomials with appropriate selection of parameters $\alpha$ and $\beta$ and  interpolation polynomials
at Chebyshev points   provide better approximation to the reciprocal  of a  %{\color{red} non-vanishing or analytic ? }
polynomial  on a cube
   than  Chebyshev polynomials  do  \cite{ncjs22}, see    Figure \ref{approximation.fig} and Table
  \ref{MaxAppErr.tab}.

%For  $\alpha, \beta>-1$, d
Define standard {\em univariate  Jacobi polynomials}  $P_n^{(\alpha, \beta)}(t), n=0, 1$ on $[-1, 1]$ by
%\begin{subequations}\label{jacobipolynomial.def}
\begin{equation*}\label{jacobipolynomiala.def}
P_0^{(\alpha, \beta)}(t)=1, \
P_1^{(\alpha, \beta)}(t)=\frac{\alpha+\beta+2}{2}t+\frac{\alpha-\beta}{2},
\end{equation*}
and $P_n^{(\alpha, \beta)}(t), n\ge 2$, by the following three-term  recurrence relation,
%For real $x$ the Jacobi polynomial can alternatively be written as by
\begin{equation*}\label{jacobipolynomialb.def}
P_n^{(\alpha, \beta)}(t)=
\big(a_{n, 1}^{(\alpha,\beta)}t- a_{n, 2}^{(\alpha,\beta)}\big)P_{n-1}^{(\alpha, \beta)}(t)-a_{n, 3}^{(\alpha,\beta)}P_{n-2}^{(\alpha, \beta)}(t),\end{equation*}
%\end{subequations}
where
$$    \begin{aligned}
        a_{n, 1}^{(\alpha,\beta)}&=\frac{(2n+\alpha+\beta-1)(2n+\alpha+\beta)}{2n(n+\alpha+\beta)},\\
        a_{n, 2}^{(\alpha,\beta)}&=\frac{(\beta^2-\alpha^2)(2n+\alpha+\beta-1)}{2n(n+\alpha+\beta)(2n+\alpha+\beta-2)},\\
        a_{n, 3}^{(\alpha,\beta)}&=\frac{(n+\alpha-1)(n+\beta-1)(2n+\alpha+\beta)}{n(n+\alpha+\beta)(2n+\alpha+\beta-2)}.
    \end{aligned}
$$
The Jacobi polynomials $P_n^{(\alpha, \beta)}, n\ge 0$, with $\alpha=\beta$ are also known as
 Gegenbauer polynomials or ultraspherical polynomials.
The Legendre polynomials $P_n$,
Chebyshev polynomials  $T_n$ and Chebyshev polynomial  of the second kind  $U_n, n\ge 0$,  %$T_n(t)=\cos (n\theta)$
%and Chebyshev polynomials of the second kind $U_n(t)=\sin ((n+1)\theta)/\sin \theta, n\ge 0$
are  Jacobi polynomials  with $\alpha=\beta=0, -1/2, 1/2$  respectively
\cite{Ismail2009, Shen2011}.
 %, where $t=\cos\theta$.

In order to
construct   polynomial  filters   to approximate the inverse of a polynomial filter of multiple graph shifts,
we next define
multivariate
  Jacobi polynomials $P_{{\bf n}; {\pmb \mu}, {\pmb \nu}}^{(\alpha, \beta)}, {\bf n}\in\mathbb{Z}_+^d$,
  and Jacobi weights $w^{(\alpha, \beta)}_{{\pmb \mu}, {\pmb \nu}}$  on the cube
$[{\pmb \mu}, {\pmb \nu}]$ by
$$P_{{\bf n};  {\pmb \mu}, {\bf v}}^{(\alpha, \beta)}({\bf t})=
\prod_{i=1}^d P_{n_i}^{(\alpha, \beta)}\left(\frac{2t_i-\mu_i-\nu_i}{\nu_i-\mu_i}\right)$$
and
$$w^{(\alpha, \beta)}_{{\pmb \mu}, {\pmb \nu}}({\bf t})=\prod_{i=1}^d
w^{(\alpha, \beta)}\left(\frac{2t_i-\mu_i-\nu_i}{\nu_i-\mu_i}\right), %{\bf t}\in [{\pmb \mu}, {\pmb \nu}],
$$
where  ${\bf t}=(t_1, \ldots, t_d) \in  [{\pmb \mu}, {\pmb \nu}]$,
${\bf n}=(n_1,\cdots, n_d)\in {\mathbb Z}_+^d$, and
$w^{(\alpha, \beta)}(t) := (1-t)^{\alpha} (1+t)^{ \beta},\ -1<t<1$.

Let $L^2( %[{\pmb \mu}, {\pmb \nu}],
w^{(\alpha, \beta)}_{{\pmb \mu}, {\pmb \nu}})$ be
 the  Hilbert space of all square-integrable functions with respect to the Jacobi weight  $ w^{(\alpha, \beta)}_{{\pmb \mu}, {\pmb \nu}}$ on $[{\pmb \mu}, {\pmb \nu}]$
 and denote
 its norm  by   $\|\cdot\|_{2,  w^{(\alpha, \beta)}_{{\pmb \mu}, {\pmb \nu}}}$.
 Following the argument in \cite{Ismail2009, Shen2011, trefethen2013} for univariate Jacobi polynomials, we can show that
multivariate Jacobi polynomials $P_{{\bf n}; {\pmb \mu}, {\pmb \nu}}^{(\alpha, \beta)}, {\bf n}\in\mathbb{Z}_+^d$,
form  a complete orthogonal system in $L^2( %[{\pmb \mu}, {\pmb \nu}],
 w^{(\alpha, \beta)}_{{\pmb \mu}, {\pmb \nu}})$
with
\begin{equation*}
\big\|P_{{\bf n}; {\pmb \mu}, {\bf v}}^{(\alpha, \beta)}\big\|_{2,  w^{(\alpha, \beta)}_{{\pmb \mu}, {\pmb \nu}}}^2
=   2^{-d} |[{\pmb \mu}, {\pmb \nu}]| \gamma_{\bf n}^{(\alpha, \beta)},
\end{equation*}
where $\Gamma(s)=\int_0^\infty t^{s-1} e^{-t} dt, s>0$, is the Gamma function, and
for
${\bf n}=(n_1,\cdots, n_d)\in\mathbb{Z}_+^d$,
$$\gamma_{\bf n}^{(\alpha, \beta)}=\prod_{i=1}^d \frac{2^{\alpha+\beta}}{2n_i+\alpha+\beta+1}\frac{\Gamma(n_i+\alpha+1)\Gamma(n_i+\beta+1)}{\Gamma(n_i+\alpha+\beta+1)\Gamma(n_i+1)}.$$
% and
%$\Gamma(s)=\int_0^\infty t^{s-1} e^{-t} dt, s>0$, is the Gamma function.

For ${\bf n}=(n_1,\cdots, n_d)\in {\mathbb Z}^d_+$, we set $\|{\bf n}\|_\infty=\sup_{1\le i\le d} |n_i|$ and define
\begin{equation} \label{Fouriercoefficient.def}
c_{\bf n}= \frac{2^d}{|[{\pmb \mu}, {\pmb \nu}]| \gamma_{\bf n}^{(\alpha, \beta)}}
\int_{[{\pmb \mu}, {\pmb \nu}]}  \frac {P_{{\bf n}; {\pmb \mu}, {\pmb \nu}}^{(\alpha, \beta)} ({\bf t})} {h({\bf t})}w^{(\alpha, \beta)}_{{\pmb \mu}, {\pmb \nu}}({\bf t}) d{\bf t}.
\end{equation}
As $1/h$ is  an analytic function on the cube $[{\pmb \mu}, {\pmb \nu}]$ by \eqref{polynomial.assump},  following the argument
in \cite[Theorem 2.2]{xiang2012}
we can show  that
the partial summation
\begin{equation}
\label{partialsum.def}
g_M^{(\alpha, \beta)}({\bf t})=\sum_{\|{\bf n}\|_\infty\le M} c_{\bf n} P_{{\bf n}; {\pmb \mu}, {\pmb \nu}}^{(\alpha, \beta)} ({\bf t}),\  M\ge 0
\end{equation}
of its  Fourier expansion converges  to $1/h$ exponentially in the uniform norm, see \cite[Theorem 8.2]{trefethen2013} for Chebyshev polynomial approximation and
\cite[Theorem 2.5]{wang2011} for Legendre polynomial approximation. This together with the boundedness of  the polynomial $h$ on the
cube  $[{\pmb \mu}, {\pmb \nu}]$ implies that
 the existence of positive constants $ D_0\in (0, \infty)$ and $r_0\in (0, 1)$ such that
\begin{equation}\label{bN.def}
    b_M^{(\alpha, \beta)}:=\sup_{{\bf t}\in [{\pmb \mu}, {\pmb \nu}]} |1-g_M^{(\alpha, \beta)}({\bf t}) h({\bf t})|\leq   D_0 r_0^M,\ M\geq 0.
\end{equation}

Shown in  Figure \ref{approximation.fig}, except the  figure on the bottom right, are
the approximation error %\st{functions}
$1-h_1(t)g_M^{(\alpha, \beta)}(t), 0\le t\le 2$, where  $g_M^{(\alpha, \beta)}, 0\le M\le 4$, are the  partial summation in \eqref{partialsum.def} to   approximate the
reciprocal $1/h_1$ of  the univariate polynomial
\begin{equation}\label{h1.def}
h_1(t) = (9/4-t)(3 + t), \ t\in [0, 2]
\end{equation}
in \cite[Eqn. (5.4)]{ncjs22}. % for some pairs of parameters $(\alpha, \beta)$.
Presented in Table \ref{MaxAppErr.tab}, except the last row,
are the maximal approximation errors measured by
$b_M^{(\alpha, \beta)}, 0\le M\le  4$.
This  demonstrates that  Jacobi polynomials have exponential approximation property \eqref{bN.def} and
also that with appropriate selection of parameters $\alpha, \beta>-1$, they have better approximation
property than Chebyshev polynomials (the Jacobi polynomials with $\alpha=\beta=-1/2$) do,  see the figure plotted on the top left of Figure \ref{approximation.fig}
and the maximal approximation errors listed in the first row of Table \ref{MaxAppErr.tab},
and  also the numerical simulations in Section \ref{circulantgraph.demo}.
% and c.f. \cite[Fig. 1]{ncjs22}.

\begin{table}[t]
		\renewcommand\arraystretch{1.2}
		\centering
		\caption{ Shown in the first seven rows are the maximal approximation error  $b_M^{(\alpha, \beta)}, 0\le M\le 4$, of Jacobi polynomial approximations
to $1/h_1$ on $[0, 2]$, while in the last row is the  maximal approximation error $\tilde b_M, 0\le M\le 4$, of Chebyshev interpolation approximation to $1/h_1$ on $[0, 2]$.
	}
		%\label{CirculantGraphICPA.Table}			
       \begin{tabular} {|c|c|c|c|c|c|c|c|c|c|}			
			\hline
			\backslashbox{$(\alpha, \beta)$}%{ME}
{M}& 0 & 1 & 2  & 3
& 4  %& 5  & 6
   \\
			\hline
			(-.5, -.5)   &  1.0463  &   0.5837  &   0.2924    &   0.1467
&   0.0728  %&   0.0367   &    0.0184
% &    0.0184    0.0092    0.0046    0.0023     0.0011    0.0006
\\			 \hline
			(.5 .5)   &
    0.7014  &  0.5904  &    0.3897  &     0.2505
     &     0.1517   %&  0.0893     & 0.0513
    %   0.0513    0.0290    0.0162    0.0089    0.0049    0.0027
    \\  \hline
			(0, 0) &
    0.7409   &  0.6153  &   0.3667   &   0.2146
    &   0.1202  %&     0.0660     &0.0357
    %   0.0357    0.0191    0.0102    0.0054    0.0028     0.0015
 \\			 \hline
	(1, 1) &     0.7140  &   0.5626  &    0.3927   &  0.2686
&   0.1720  % &    0.1066  &0.0643
%  0.0643    0.0380    0.0220    0.0126    0.0072    0.0040
\\			 \hline
			(-.5, .5) &    1.8612 &     1.8855  &    1.3522    &    0.8937
&    0.5534   %&   0.3304  & 0.1920
%  0.1920    0.1094    0.0614    0.0340    0.0187    0.0102
 \\			 \hline
(.5, -.5) &    0.7720  &    0.5603  &    0.3563   &    0.2184
&    0.1289  %&   0.0744 &0.0422
%   0.0422    0.0236    0.0130    0.0071    0.0039    0.0021
 \\			 \hline
(0, -.5) &
    0.7356 &    0.4760  &    0.2749  &    0.1548
    &  0.0850  %&    0.0460     & 0.0246
    %  0.0246    0.0131    0.0069    0.0036    0.0019    0.0010
% \\			 \hline
%(0, -.8)  &    0.7787  &   0.4993   &  0.2927   &    0.1512   &    0.0879  &   0.0470
%%   0.0451    0.0449    0.0456    0.0465    0.0472    0.0480
  \\			 \hline
  {ChebyInt} &
  0.7500   &  0.4497 &     0.2342  &     0.1186
  &     0.0595  %&    0.0298    &0.0149
  %   0.0074    0.0037    0.0019    0.0009
    \\ \hline
		\end{tabular}
		\label{MaxAppErr.tab}
	\end{table}

%
%    0.5329    0.4342    0.2116    0.1309    0.0678    0.0378    0.0198    0.0106    0.0056    0.0029    0.0015
%
%  Columns 12 through 21
%
%    0.0008    0.0004    0.0002
%
%         0.6250    0.4389    0.2093    0.1114    0.0544    0.0276    0.0137    0.0069    0.0034    0.0017    0.0009
%
%  Columns 12 through 13
%
%    0.0004    0.0002

% Chebyshev Interpolation maximal error Columns 1 through 11
%
%    0.7500    0.4497    0.2342    0.1186    0.0595    0.0298    0.0149    0.0074    0.0037    0.0019    0.0009
%
%  Columns 12 through 13
%
%    0.0005    0.0002

Another excellent method of approximating the reciprocal  % $1/h$
 of the polynomial $h$ on the cube $[{\pmb \mu}, {\bf \pmb \nu}]$
is  polynomial  interpolation
 \begin{equation}\label{Chebyshevinterpolation.def}
 C_M({\bf t})=\sum_{\|{\bf n}\|_\infty\le M} d_{\bf n}{\bf t}^{\bf n}\end{equation}
  at rescaled Chebyshev points
 ${\bf t}_{{\bf j}; {\pmb \mu}, {\pmb \nu} }=
 ({t}_{j_1, M},  \ldots, {t}_{j_d, M})$, i.e.,
 \begin{equation}
\label{chebyshevinterpolation.def}
C_M({\bf t}_{{\bf j}; {\pmb \mu}, {\pmb \nu} })= 1/h( {\bf t}_{{\bf j}; {\pmb \mu}, {\pmb \nu} }),
\end{equation}
%$C_M({\bf t}_{{\bf j}; {\pmb \mu}, {\pmb \nu} })= 1/h( {\bf t}_{{\bf j}; {\pmb \mu}, {\pmb \nu} })$,
 where
 $${ t}_{j_k, M}= \frac{\nu_k+\mu_k}{2}+ \frac{\nu_k-\mu_k}{2}
  \cos \frac{(j_k-1/2)\pi}{M+1}$$
  for $1\le j_k\le M+1, 1\le k\le d$.
% , \ldots,
%  \frac{\nu_d-\mu_d}{2} \cos \frac{(j_d-1/2)\pi}{N+1}\right),$$
% where
Recall that the  Lebesgue constant for the above polynomial interpolation at rescaled Chebyshev points
is of the order $(\ln (M+2))^d$. This together with
the exponential convergence
of Chebyshev polynomial approximation, see \cite[Theorem 8.2]{trefethen2013} and \cite[Theorem 2.2]{xiang2012}, implies that
\begin{equation}\label{exponentialapproximationerror.interpolation}
\tilde b_M:=\sup_{{\bf t}\in  [{\pmb \mu}, {\pmb \nu}]}
|1-h({\bf t})C_M({\bf t})|\le D_1 r_1^M, \ M\ge 0,
\end{equation}
for some  positive constants
  $D_1\in (0, \infty)$ and $r_1\in (0, 1)$.
  Shown in
   the bottom right of Figure \ref{approximation.fig}
   is our numerical demonstration to the  above    approximation
  property of the Chebyshev interpolation polynomial $C_M$, ChebyInt for abbreviation,  to the function $1/h_1$,  see bottom row of Table
  \ref{MaxAppErr.tab}
      for the  maximal approximation error $\tilde b_M, 0\le M\le 4$, in \eqref{exponentialapproximationerror.interpolation}
      and
      also the numerical simulations in Section \ref{circulantgraph.demo}.

\section{Polynomial approximation algorithm for inverse filtering}
\label{Jacobiapproximation.section}

Let  ${\mathbf S}_1, \ldots, {\bf S}_d$ be commutative graph shifts   whose joint spectrum $\Lambda$ in \eqref{jointspectrum.def} is contained in
a cube $[{\pmb \mu}, {\pmb \nu}]$, i.e., \eqref {jointspectralcubic.def} holds.
 The  joint spectrum  $\Lambda$ of commutative graph shifts
${\bf S}_1, \ldots, {\bf S}_d$
 plays a critical role in  \cite{ncjs22}
  to construct  optimal/Chebyshev polynomial  approximation  to the inverse  of a polynomial filter.
In  this section, based on the exponential approximation property of Jacobi polynomials and Chebyshev interpolation polynomials to the reciprocal of a nonvanishing multivariate polynomial,
 we propose an iterative Jacobi polynomial approximation algorithm  and
 Chebyshev interpolation approximation algorithm
to implement the inverse filtering procedure associated with a polynomial graph filter at the vertex level with one-hop communication. %  ${\bf S}_1, \ldots, {\bf S}_d$.

Let  $\alpha, \beta>-1$,
 $h$ be a multivariate polynomial   satisfying
\eqref{polynomial.assump}, and
let $g_M^{(\alpha, \beta)}$ and $C_M, M\ge 0$, be the  Jacobi polynomial approximation and Chebyshev interpolation polynomial approximation to $1/h$
in \eqref{partialsum.def} and \eqref{chebyshevinterpolation.def} respectively.
 Set ${\bf H}=h({\bf S}_1, \ldots, {\bf S}_d)$,
${\bf G}_M^{(\alpha, \beta)}= g_M^{(\alpha, \beta)}({\bf S}_1, \ldots, {\bf S}_d)$  and  ${\bf C}_M=C_M({\bf S}_1, \ldots, {\bf S}_d), M\ge 0$.
By the  spectral assumption \eqref{jointspectralcubic.def}, % on commutative graph shifts ${\bf S}_1, \ldots, {\bf S}_d$,
%we have that
the spectral radii of ${\bf I}-{\bf G}_M^{(\alpha, \beta)}{\bf H}$ and
${\bf I}-{\bf C}_M{\bf H}$
 are bounded by $b_M^{(\alpha, \beta)}$ in \eqref{bN.def}
 and $\tilde b_M$  in \eqref{exponentialapproximationerror.interpolation}
respectively, i.e.,
\begin{equation}
\rho({\bf I}-{\bf G}_M^{(\alpha, \beta)}{\bf H})\leq b^{(\alpha, \beta)}_M\
{\rm and}
\
\rho({\bf I}-{\bf C}_M{\bf H})\leq \tilde b_M,
 \ M\ge 0.
\end{equation}
Therefore with appropriate selection of the polynomial degree $M$,
 applying the arguments used in \cite[Theorem 3.1]{ncjs22}, we obtain the exponential convergence of  the following iterative algorithm for inverse filtering,
\begin{equation} \label{jacobiapproximation.def}  %\label{jacobiapproximation.eqa}
\left\{\begin{array}{l}
{\bf e}^{(m)} = {\bf H} {\bf  x}^{(m-1)} -  {\bf y}\\
{\bf x}^{(m)} = {\bf x}^{(m-1)} -{\bf G}_M {\bf e}^{(m)}, \ m\ge 1
\end{array}
\right.
\end{equation}
with  arbitrary initials
${\bf x}^{(0)}$,
where  ${\bf G}_M$ is either ${\bf G}_M^{(\alpha, \beta)}$ or ${\bf C}_M$, and
 the input ${\bf y}$ of the inverse filtering procedure
   is obtained via the filtering procedure \eqref{filtering.def}.

\begin{thm}\label{exponentialconvergence.thm}
Let ${\mathbf S}_1, \ldots, {\bf S}_d$ be commutative graph shifts satisfying
\eqref{jointspectralcubic.def},
  $h$ be a multivariate polynomial satisfying \eqref{polynomial.assump},
and let  $b_M^{(\alpha, \beta)}$
 and $\tilde b_M$ be given in \eqref{bN.def} and \eqref{exponentialapproximationerror.interpolation} respectively.
If
\begin{equation}
b_M^{(\alpha, \beta)}<1 \ \ ({\rm resp.} \ \tilde b_M<1),\end{equation}
 then for any  input ${\bf y}$, the sequence
${\bf x}^{(m)}, m\ge 0$, in the iterative algorithm \eqref{jacobiapproximation.def}  with ${\bf G}_M={\bf G}_M^{(\alpha, \beta)}$ (resp.
  ${\bf G}_M={\bf C}_M$)
converges to  the output ${\bf H}^{-1}{\bf y}$ of the inverse filtering procedure \eqref{inverseprocedure} exponentially. In particular,
there exist constants $C\in (0, \infty)$ and $r\in (\rho({\bf I}-{\bf G}_M^{(\alpha, \beta)}{\bf H}), 1)$  (resp. $r\in (\rho({\bf I}-{\bf C}_M{\bf H}), 1)$) such that
\begin{equation}
\| {\bf x}^{(m)}- {\bf H}^{-1}{\bf y}\|_2 \le C \|{\bf y}\|_2 r^m, \ m\ge 0.
\end{equation}
\end{thm}

 We call the
  algorithm \eqref{jacobiapproximation.def}  with ${\bf G}_M={\bf G}_M^{(\alpha, \beta)}$
  as  {\em   Jacobi polynomial approximation algorithm},  JPA$(\alpha, \beta)$ for abbreviation,
  and  the iterative
  algorithm \eqref{jacobiapproximation.def}  with ${\bf G}_M={\bf C}_M$
  as  {\em  Chebyshev interpolation polynomial approximation algorithm}, CIPA for abbreviation.
By  Theorem \ref{exponentialconvergence.thm}, the exponential convergence rates of
the  JPA$(\alpha, \beta)$
and    CIPA are
$\rho({\bf I}-{\bf G}_M^{(\alpha, \beta)} {\bf H})$ and
$\rho({\bf I}-{\bf C}_M {\bf H})$ respectively.
 In addition to the  exponential convergence, % shown in Theorem \ref{exponentialconvergence.thm},
each iteration in the JPA$(\alpha, \beta)$ and  CIPA  contains essentially two filtering procedures associated
with polynomial filters  ${\bf G}_M$  and ${\bf H}$,
and hence it can be implemented at the  vertex level with one-hop communication,  see
 \cite[Algorithm 4] {ncjs22}.
Therefore the JPA$(\alpha, \beta)$ and  CIPA algorithms
%the iterative Jacobi polynomial approximation algorithm and  the iterative  Chebyshev interpolation polynomial approximation algorithm
   can be implemented  on a network with each agent equipped with limited storage and data processing ability,  and one-hop communication subsystem.
  More importantly,  the memory, computational cost and communication expense for
each agent of the network are {\bf independent} on the size of the whole network.

\begin{remark} {\rm  We remark that the  JPA$(\alpha, \beta)$  with $\alpha=\beta=-1/2$ was introduced in
\cite{ncjs22} as iterative Chebyshev polynomial approximation algorithm.
 For a positive definite polynomial filter ${\bf H}$,
 replacing the approximation filter ${\bf G}_M$
in  the  quasi-Newton algorithm  \eqref{jacobiapproximation.def} by  $\gamma_{\rm opt} {\bf I}$, we obtain the
traditional gradient descent method
\begin{equation} \label{gd0.def}  %\label{jacobiapproximation.eqa}
{\bf x}^{(m)} = {\bf x}^{(m-1)} -\gamma_{\rm opt} ({\bf H}   {\bf  x}^{(m-1)} -  {\bf y}), \ m\ge 1
\end{equation}
 with the optimal step size $\gamma_{\rm opt}={2}/{(\lambda_{\min}({\bf H})+\lambda_{\rm max}({\bf H})})$, where $\lambda_{\max} ({\bf H})$ and $\lambda_{\min} ({\bf H})$ are the maximal and minimal eigenvalue of the matrix ${\bf H}$ respectively \cite{jiang19, Leus17, Waheed18, Shi15, sihengTV15, Shuman18, isufi19}.
 Numerical comparisons
with the JPA$(\alpha, \beta)$ and  CIPA algorithms
to implement inverse filtering on circulant graphs will be given in Section \ref{circulantgraph.demo}.
%\ref{simulation.section}.
}
\end{remark}

\section{Wiener filters for  stationary graph signals}
\label{stochasticwienerfilter.section}

Let ${\mathbf S}_1, \ldots, {\mathbf S}_d$ be real commutative  symmetric graph shifts on a simple graph ${\mathcal G}=(V, E)$ of order $N\ge 1$
and assume that
 their joint spectrum is contained in some cube
$[{\pmb \mu}, {\pmb \nu}]$,  i.e., \eqref{jointspectralcubic.def} holds.
In this section, we consider the scenario that
the filtering procedure \eqref{filtering.def} has the  filter
 \begin{subequations}\label{wiener2.eq}
  \begin{equation} \label{wiener2.eqb}
  {\bf H}=h({\mathbf S}_1, \ldots, {\bf S}_d)
  \end{equation}
 being a
  polynomial filter of ${\bf S}_1, \ldots, {\bf S}_d$,  %for some multivariate polynomial $h$,
 the inputs
${\bf x}$ %=(x(i))_{i\in V}$
are
{\em stationary} signals with the correlation  matrix % ${\bf R}={\mathbb E} ({\bf x} {\bf x}^T)$
\begin{equation}  \label{wiener2.eq1}
{\bf R}= r({\mathbf S}_1, \ldots, {\mathbf S}_d)
\end{equation}
being a polynomial of graph shifts ${\bf S}_1, \ldots, {\mathbf S}_d$
(\cite{perraudin17, segarrat2017, yagan2020}), and
 the  outputs
 \begin{equation}\label{wiener2.eq2}
{\bf y}={\bf H} {\bf x}+{\pmb \epsilon}\end{equation}
 are corrupted by some random noise ${\pmb \epsilon}$
being independent with the input signal ${\bf x}$, and  having zero  mean   and
 covariance matrix  ${\bf G}$ to be
  a polynomial of graph shifts ${\bf S}_1, \ldots, {\mathbf S}_d$, i.e.,
\begin{equation} \label{wiener2.eq3} {\mathbb E}{\pmb \epsilon}={\bf 0},  \  {\mathbb E}{\pmb \epsilon}{\bf x}^T={\bf 0}
 \ {\rm and}\  {\bf G}=g({\bf S}_1, \ldots, {\bf S}_d)
\end{equation}
\end{subequations}
for some multivariate polynomial $g$. In this section, we find the optimal reconstruction filter ${\bf W}_{\rm mse}$
% \st{we find the stochastic Wiener filter ${\bf W}_{\rm mse}$ which is the optimal reconstruction filter}
with respect to the stochastic mean squared error
$F_{{\rm mse},P, {\bf K}}$ in \eqref{mse.objectivefunction}, and  we propose a distributed algorithm to
implement the stochastic Wiener filtering procedure ${\bf y}\longmapsto {\bf W}_{\rm mse} {\bf y}$ at the vertex
level with one-hop communication. %\st{with data exchanging with adjacent vertices only}.
In this section, we also consider   optimal unbiased reconstruction filters for the scenario that the input signals
${\bf x}$  are  {\em  wide-band stationary}, i.e.,
\begin{equation}\label{wiener2.eq4}
{\mathbb E}{\bf x}=c {\bf 1} \ \ {\rm and} \ \ {\mathbb E}({\bf x}-{\mathbb E}{\bf x}) ({\bf x}-{\mathbb E}({\bf x}))^T
=\widetilde {\bf R}= \tilde r({\bf S}_1, \ldots, {\bf S}_d),
\end{equation}
for some  $0\ne c\in {\mathbb R}$  %{\color{red} ${\bf 1}$ has been defined in the notation section} \st{ is the column vector with all entries taking value one},
and  some multivariate polynomial  $\tilde r$,
The concept of (wide-band) stationary signals was introduced in \cite[Definition 3]{perraudin17} in which the
graph Laplacian is used as the graph shift.

For a  probability measure  $P=(p(i))_{i\in V}$ on the graph ${\mathcal G}$
and a regularization matrix ${\bf K}$, we define
 the {\em stochastic mean squared error}
of  a reconstruction filter ${\bf W}$
by
\begin{equation}\label{mse.objectivefunction}
F_{{\rm mse}, P, {\bf K}}({\bf W}) %& \hskip-0.08in =  & \hskip-0.08in {\mathbb E}  \sum_{i\in V} p(i)
% | ({\bf W} {\bf y})(i)-{ x}(i)|^2 +  {\bf y}^T {\bf W}^T {\bf K}  {\bf W} {\bf y}\nonumber\\
% & \hskip-0.08in =  & \hskip-0.08in {\mathbb E}
={\mathbb E}  ({\bf W}{\bf y}-{\bf x})^T {\bf P} ({\bf W}{\bf y}-{\bf x}) +  {\bf y}^T {\bf W}^T {\bf K}  {\bf W} {\bf y},
\end{equation}
where ${\bf P}$ is the diagonal matrix with diagonal entries $p(i), i\in V$. The stochastic mean squared error  $F_{{\rm mse}, P, {\bf K}}({\bf W})$ in \eqref{mse.objectivefunction}
 contains
the regularization term ${\mathbb E} {\bf y}^T {\bf W}^T {\bf K}  {\bf W} {\bf y}$
and the fidelity term ${\mathbb E} ({\bf W}{\bf y}-{\bf x})^T {\bf P} ({\bf W}{\bf y}-{\bf x})= \sum_{i\in V} p(i)
{\mathbb E}  | ({\bf W} {\bf y})(i)-{ x}(i)|^2$. It is discussed in \cite{perraudin17} for the case that the filter ${\bf H}$, the covariance ${\bf G}$ of noises and the regularizer ${\bf K}$
are polynomials of  the graph Laplacian ${\bf L}$, and that
the probability measure $P$ is
the uniform probability measure $P_U$, i.e., $p_U(i)=1/N, i\in V$.
In the following theorem, we provide an explicit solution to the minimization
$\min_{\bf W} F_{{\rm mse}, P, {\bf K} }({\bf W})$, see Appendix \ref{wienerfiltermsd.prof} for the proof.

\begin{thm} \label{wienerfiltermse.thm}
Let  the filter  ${\bf H}$, the input signal ${\bf x}$,
the  noisy  output signal  ${\bf y}$ and additive noise ${\pmb \epsilon}$ be as in \eqref{wiener2.eq}, and let the stochastic mean squared error $F_{{\rm mes}, P, {\bf K}}$
be as in \eqref{mse.objectivefunction}. Assume that
${\bf H}{\bf R}{\bf H}^T+{\bf G}$  and ${\bf P}+{\bf K}$ are strictly positive definite,
and define
\begin{equation}\label{wienerfiltermse.eq2}
{\bf W}_{{\rm mse}}= ({\bf P}+{\bf K})^{-1}{\bf P}  {\bf R}{\bf H}^T \big( {\bf H} {\bf R}{\bf H}^T+{\bf G})^{-1}.
\end{equation} Then
${\bf W}_{{\rm mse}}$ is the unique minimizer of the minimization problem
 \begin{equation} \label{wienerfiltermse.eq0}
 {\bf W}_{{\rm mse}}=\arg \min_{\bf W} F_{{\rm mse}, P, {\bf K} }({\bf W}),\end{equation}
 and
\begin{equation} \label{wienerfiltermse.eq1}
  F_{{\rm mse}, P, {\bf K} }({\bf W}_{\rm mse})
  =
 {\rm tr}\big ({\bf P}({\bf I}- {\bf W}_{\rm mse} {\bf H}\big){\bf R}\big).
 %{\bf R} {\bf H}^T ({\bf H}{\bf R} {\bf H}^T +{\bf G})^{-1} {\bf H}{\bf R}\right)\nonumber\\
 %& & - {\rm tr}\left ({\bf K}({\bf P}+{\bf K})^{-1} {\bf P} ({\bf P}{\bf K})^{-1} {\bf P} {\bf R} {\bf H}^T ({\bf H}{\bf R} {\bf H}^T +{\bf G})^{-1} {\bf H}{\bf R}\right)
\end{equation}
\end{thm}

We call the optimal reconstruction filter   ${\bf W}_{\rm mse}$ in \eqref{wienerfiltermse.eq2} as
the  {\em  stochastic Wiener filter}.
For the case that   the stochastic mean squared error does not take the regularization term  into account, i.e., ${\bf K}={\bf 0}$,
we obtain from \eqref{wienerfiltermse.eq2}
 that the corresponding stochastic Wiener filter ${\bf W}_{\rm mse}$ becomes
 \begin{equation}\label{wmse0.def}
 {\bf W}_{\rm mse}^0= {\bf R}{\bf H}^T \big( {\bf H} {\bf R}{\bf H}^T+{\bf G})^{-1},\end{equation}
  which is independent of the probability measure  $P=(p(i))_{i\in V}$ on the graph ${\mathcal G}$.
  If we further assume that the probability measure $P$ is
the uniform probability measure $P_U$
 and the input signals ${\bf x}$ are i.i.d with mean zero and variance $\delta_1$,
the stochastic Wiener filter becomes
$${\bf W}_{\rm mse}^0= \delta_1^2 {\bf H}^T (\delta_1^2 {\bf H}{\bf H}^T+{\bf G})^{-1}$$
 and
the corresponding
stochastic mean squared error  is given by
\begin{equation}\label{fmsepu}
F_{{\rm mes}, P_U}({\bf W}_{\rm mse}^0)=
\frac{\delta_1^2}{N} {\rm tr}  \left ( (\delta_1^2 {\bf H}{\bf H}^T +{\bf G})^{-1} {\bf G}\right),
\end{equation}
cf. \eqref{wienerfilterworsecase.eq2} and \eqref{wienerfilterworsecase.error2}, and \cite[Eqn.  16]{perraudin17}.

Denote the reconstructed signal via the stochastic Wiener filter ${\bf W}_{\rm mse}$ by
\begin{equation}\label{stochasticestimator}
 {\bf x}_{\rm mse}={\bf W}_{\rm mse} {\bf y},\end{equation}
where ${\bf y}$ is given in \eqref{wiener2.eq2}.
The above estimator  via stochastic Wiener filter ${\bf W}_{\rm mse}$ is {\bf biased} in general. For the case that
${\bf G}, {\bf H}, {\bf K}$ and ${\bf R}$ are polynomials of commutative symmetric graph shifts ${\mathbf S}_1, \ldots, {\mathbf S}_d$,
one may verify that
matrices ${\bf H}^T, {\bf H}, {\bf G}, {\bf R}, {\bf K}$ are commutative,  and
 \begin{eqnarray}
{\mathbb E}({\bf x}-{\bf x}_{\rm mse}) & \hskip-0.08in = & \hskip-0.08in ({\bf P}+{\bf K})^{-1}  \big( {\bf H} {\bf R}{\bf H}^T+{\bf G})^{-1}  {\bf R}{\bf H}^T {\bf H}{\bf K} {\mathbb E}{\bf x}\nonumber\\
%& \hskip-0.08in - & \hskip-0.08in ({\bf R}{\bf H}^T \big( {\bf H} {\bf R}{\bf H}^T+{\bf G})^{-1} {\bf H}-{\bf I}){\mathbb E}{\bf x}\nonumber\\
%\\
& \hskip-0.08in  & \hskip-0.08in +
\big( {\bf H} {\bf R}{\bf H}^T+{\bf G})^{-1} {\bf G} {\mathbb E}{\bf x}.
\end{eqnarray}
Therefore the estimator \eqref{stochasticestimator} is {\bf unbiased}  if
\begin{equation}\label{stochasticestimator.unbiased}
{\mathbf K}{\mathbb E}{\bf x}={\mathbf G}{\mathbb E}{\bf x}={\bf 0}.
\end{equation}

\begin{remark}\label{stochasticwiener.remark}
{\rm By  \eqref{wienerfiltermse.eq2} and
\eqref{wmse0.def},   the reconstructed signal ${\bf x}_{\rm mse}$ in \eqref{stochasticestimator} can be obtained in two steps,
\begin{subequations} \label{stochasticrecovery.eq}
\begin{equation} \label{stochasticrecovery.eqa}
{\bf w}= {\bf W}_{\rm mse}^0{\bf y}= {\bf R} {\bf H}^T \big({\bf H} {\bf  R} {\bf H}^T +{\bf G}\big)^{-1} {\bf y},
\end{equation}
and
\begin{equation} \label{stochasticrecovery.eqb}
%\left\{
%\begin{array}{l}
%{\bf z}_1 = \big({\bf H} {\bf  R} {\bf H}^T +{\bf G}\big)^{-1} {\bf y},\\
% {\bf z}_2= {\bf R} {\bf H}^T {\bf z}_1,\\
%{\bf z}_3= {\bf P}^{1/2} {\bf z}_2,\\
%{\bf z}_4=\big({\bf I}+{\bf P}^{-1/2} {\bf K}{\bf P}^{-1/2})^{-1} {\bf z}_3,\\
 {\bf x}_{\rm mse}%= ({\bf P}+{\bf K})^{-1} {\bf P} {\bf w}
 = {\bf P}^{-1/2}  \big({\bf I}+{\bf P}^{-1/2} {\bf K}{\bf P}^{-1/2})^{-1} {\bf P}^{1/2}  {\bf w},
%\end{array}
%\right.
\end{equation}
\end{subequations}
%and applying of a filtering procedure
%\begin{equation} \label{stochasticrecovery.eqa} \tilde {\bf x}= {\bf R} {\bf H}^T {\bf z}.
%\end{equation}
where the first step \eqref{stochasticrecovery.eqa} is the Wiener filtering procedure
 without the regularization term  taken into account, and the second step \eqref{stochasticrecovery.eqb} is the solution of the
 following Tikhonov regularization problem,
\begin{equation}\label{tihhonov.eq0} {\bf x}_{\rm mse}=\arg\min_{\bf x}\ ({\bf x}-{\bf w})^T {\bf P} ({\bf x}-{\bf w})+{\bf x}^T {\bf K} {\bf x}.\end{equation}
}
\end{remark}

By   symmetry and commutativity  assumptions on the graph shifts ${\mathbf S}_1, \ldots, {\mathbf S}_d$, and the
 polynomial assumptions \eqref{wiener2.eqb}, \eqref{wiener2.eq1} and \eqref{wiener2.eq3}, the Wiener filter ${\bf W}_{\rm mse}^0$ in
\eqref{wmse0.def} is the product of a polynomial filter
${\bf R} {\bf H}^T=(hr)({\mathbf S}_1, \ldots, {\mathbf S}_d)$
and the inverse of another polynomial filter
 $ {\bf H}{\bf R} {\bf H}^T +{\bf G}= (h^2 r +g)({\mathbf S}_1, \ldots, {\bf S}_d)$.
 Set ${\bf z}_1 = ({\bf H} {\bf  R} {\bf H}^T +{\bf G})^{-1} {\bf y}$.
Therefore using \cite[Algorithms 1 and  2]{ncjs22}, the filtering procedure
${\bf w}= {\bf R} {\bf H}^T {\bf z}_1$ can  be implemented at the vertex level with one-hop communication.  %  \st{with  data exchanging with adjacent vertices only.
Also we observe that  the Jacobi  polynomial approximation algorithm and Chebyshev interpolation  polynomial approximation algorithm  in Section \ref{Jacobiapproximation.section}
 can be applied to the inverse filtering procedure ${\bf y}\longmapsto {\bf z}_1$, % = \big({\bf H} {\bf  R} {\bf H}^T +{\bf G}\big)^{-1} {\bf y}$
   when
 \begin{equation}\label{stochasticwienerfilter.filtercondition}
   h^2({\bf t}) r({\bf t})+g({\bf t})>0 \ {\rm for \ all} \ {\bf t}\in [{\pmb \mu}, {\pmb \nu}],\end{equation}
see Part I of Algorithm \ref{Wiener2.algorithm} for the implementation
of the Wiener filtering procedure  \eqref{stochasticrecovery.eqa} without regularization at the vertex level. %t a vertex $i\in V$.

\smallskip

Set
${\bf z}_2= {\bf P}^{1/2} {\bf w}$ and $
{\bf z}_3=\big({\bf I}+{\bf P}^{-1/2} {\bf K}{\bf P}^{-1/2})^{-1} {\bf z}_2$.
As ${\bf P}$ is a diagonal matrix, the rescaling procedure ${\bf z}_2= {\bf P}^{1/2} {\bf w}$ and
${\bf x}_{\rm mse}={\bf P}^{-1/2} {\bf z}_3$ can be implemented  at the vertex level.
Then  it remains to find a distributed algorithm to implement the inverse filtering procedure
\begin{equation}\label{inversez3}
{\bf z}_3=\big({\bf I}+{\bf P}^{-1/2} {\bf K}{\bf P}^{-1/2})^{-1} {\bf z}_2\end{equation}
 at the vertex level.
 As ${\bf P}^{-1/2}$ {\bf  may  not commutate} with the graph shifts ${\bf S}_1, \ldots, {\bf S}_d$, the filter  ${\bf I}+{\bf P}^{-1/2} {\bf K}{\bf P}^{-1/2}$ is not necessarily
  a polynomial filter of some commutative graph shifts even if ${\bf K}=k({\bf S}_1, \ldots, {\bf S}_d)$ is,  hence
   the  polynomial approximation algorithm  proposed in Section \ref{Jacobiapproximation.section}
 {\bf does  not apply} to the  above inverse filtering procedure  directly.

Next we propose a {\em novel} exponentially convergent algorithm to implement
the inverse filtering procedure  \eqref{inversez3}
%${\bf z}_3=\big({\bf I}+{\bf P}^{-1/2} {\bf K}{\bf P}^{-1/2})^{-1} {\bf z}_2$
at the vertex level
when  the positive semidefinite regularization matrix
${\bf K}=k({\bf S}_1, \ldots, {\bf S}_d)$ is a polynomial of graph shifts
${\bf S}_1, \ldots, {\bf S}_d$.
Set
$$K=\sup_{{\bf t}\in [{\pmb \mu}, {\pmb \nu}]} k({\bf t}) \   {\rm and} \  p_{\min}=\min_{i\in V} p(i).$$
 Then one may verify that
\begin{equation}\label{PKinverse.eq0}
 {\bf I}\preceq {\bf I}+{\bf P}^{-1/2} {\bf K}{\bf P}^{-1/2} \preceq \frac{ K+p_{\min}}{p_{\min}} {\bf I},\end{equation}
 where  for symmetric matrices ${\bf A}$ and ${\bf B}$, we use ${\bf A}\preceq {\bf B}$  to denote the positive semidefiniteness of ${\bf B}-{\bf A}$.
Applying Neumann series expansion $ (1-t)^{-1}=\sum_{n=0}^\infty t^n$ with $t$ replaced by
${\bf I}- \frac{p_{\min}} { K+p_{\min}} ({\bf I}+{\bf P}^{-1/2} {\bf K}{\bf P}^{-1/2})$, we obtain
\begin{eqnarray*}%  \label{PKinverse.eq1}
&\hskip-0.08in  &  \hskip-0.08in \big({\bf I}+{\bf P}^{-1/2} {\bf K}{\bf P}^{-1/2})^{-1}\nonumber\\
% =  \frac{d_{\min}} { K+d_{\min}}  \left({\bf I}- \frac{d_{\min}} { K+d_{\min}} ({\bf I}+{\bf P}^{-1/2} {\bf K}{\bf P}^{-1/2}\right)^{-1}\\
&\hskip-0.08in  =  &  \hskip-0.08in \frac{p_{\min}} { K+p_{\min}} \sum_{n=0}^\infty \left(\frac{ K {\bf I}- p_{\min}{\bf P}^{-1/2} {\bf K}{\bf P}^{-1/2}} { K+p_{\min}} \right)^n.\qquad
\end{eqnarray*}
Therefore the sequence  ${\bf w}_m, m\ge 0$,  defined by
\begin{eqnarray}
\label{PKinverse.eq2}
{\bf w}_{m+1} &\hskip-0.08in  = &  \hskip-0.08in \frac{p_{\min}}{K+p_{\min}} {\bf w}_0+ \frac{K}{K+p_{\min}} {\bf w}_m\nonumber \\
& &\hskip-0.08in  -\frac{p_{\min}}{K+p_{\min}} {\bf P}^{-1/2} {\bf K}{\bf P}^{-1/2}{\bf w}_m, \ m\ge 0
\quad \end{eqnarray}
with initial  ${\bf w}_0={\bf z}_2$
 converges to ${\bf z}_3$ exponentially, since
 \begin{eqnarray*}
  & \hskip-0.08in  & \hskip-0.08in  \|{\bf w}_m- {\bf z}_3\|_2\\
  & \hskip-0.08in = & \hskip-0.08in \frac{p_{\min}} { K+p_{\min}} \left\|\sum_{n=m+1}^\infty \left(\frac{ K {\bf I}- p_{\min}{\bf P}^{-1/2} {\bf K}{\bf P}^{-1/2}} { K+p_{\min}} \right)^n {\bf z}_2\right\|_2\\
  & \hskip-0.08in \le & \hskip-0.08in \frac{p_{\min}\|{\bf z}_2\|_2} { K+p_{\min}} \sum_{n=m+1}^\infty \left\|\frac{ K {\bf I}- p_{\min}{\bf P}^{-1/2} {\bf K}{\bf P}^{-1/2}} { K+p_{\min}} \right\|^n \\
 & \hskip-0.08in \le  & \hskip-0.08in  \frac{p_{\min}\|{\bf z}_2\|_2} { K+p_{\min}} \sum_{n=m+1}^\infty \left(\frac{K} { K+p_{\min}}\right)^n\nonumber\\
 & \hskip-0.08in = & \hskip-0.08in
 \left(\frac{K} { K+p_{\min}}\right)^{m+1} \|{\bf z}_2\|_2,\  m\ge 1,
  \end{eqnarray*}
  where the last inequality follows from \eqref{PKinverse.eq0}.
 More importantly, each iteration in the algorithm to implement the inverse filtering procedure
 \eqref{inversez3}
 % ${\bf z}_3=\big({\bf I}+{\bf P}^{-1/2} {\bf K}{\bf P}^{-1/2})^{-1} {\bf z}_2$
  contains mainly two rescaling procedure and  a filter procedure associated with the polynomial filter ${\bf K}$ which can be implemented by
   \cite[Algorithms 1 and 2]{ncjs22}.
   Hence  the  regularization procedure
   \eqref{stochasticrecovery.eqb}
can  be implemented at the vertex level with one-hop communication, see Part 2 of Algorithm \ref{Wiener2.algorithm}.

      \begin{algorithm}[t]
\caption{Polynomial approximation algorithm  to implement the Wiener filtering procedure ${\bf x}_{\rm mse}={\bf W}_{\rm mse} {\bf y}$
  at a vertex $i\in V$. }
\label{Wiener2.algorithm}
\begin{algorithmic}  %[1]

\STATE {\em Inputs}: Polynomial coefficients of polynomial filters ${\bf H}, {\bf G}, {\bf K}, {\bf R}$ and
${\bf G}_M$ (either Jacobi polynomial approximation filter  ${\bf G}_M^{(\alpha, \beta)}$ or Chebyshev interpolation approximation filter ${\bf C}_M$
to the inverse filter $({\bf H}^2 {\bf  R} +{\bf G})^{-1}$),  entries $S_k(i,j), j\in {\mathcal N}_i$ in the $i$-th row of the shifts ${\bf S}_k, 1\le k\le d$,
the  value $y(i)$  of the input signal ${\bf y}=(y(i))_{i\in V}$ at the vertex $i$, the probability $p(i)$ at the vertex $i$,  and numbers $L_1$ and  $L_2$ of the first and second iteration.

%\STATE {\bf Operation}: Evaluate $m_k=\mu(B(k, r))$, compute ${\bf F}_k= {\bf H}_{0,k}^T{\bf H}_{0,k}+ {\bf H}_{1,k}^T{\bf H}_{1,k}$,
% find its inverse  $({\bf F}_k)^{-1}$, and then compute $ {\bf G}^L_{l; k}:=({\bf F}_k)^{-1} {\bf H}_{l,k}^T, l=0, 1$.

\STATE {\bf Part I}: \ Implementation of the Wiener filtering procedure \eqref{stochasticrecovery.eqa} at the vertex $i$

\STATE {\em Pre-processing}:\  Find the polynomial coefficients of polynomial filters ${\bf H}^2 {\bf  R}  +{\bf G}$ and ${\bf R}{\bf H}$.

\STATE {\em Initialization}:  \ $n=0$ and zero initial  $x^{(0)}(i)=0$.

\STATE {\em   Iteration}: \ Use
\cite[Algorithms 1 and  2]{ncjs22}
to implement the filtering procedures
${\bf e}^{(m)}=({\bf H} {\bf  R} {\bf H}^T +{\bf G}) {\bf x}^{(m-1)}-{\bf y}$ and
${\bf x}^{(m)}= {\bf x}^{(m-1)}-{\bf G}_M {\bf e}^{(m)}, 0\le m\le L_1$
at the vertex $i$.

\STATE {\em  Output of the iteration}: \
 Denote the output of the $L_1$-th iteration  by
$ z_1^{(L_1)}(i)$, which is  the approximate value of the output data of the inverse filtering procedure
${\bf z}_1 = ({\bf H}^2 {\bf  R}  +{\bf G})^{-1} {\bf y}$ at the vertex $i$.

\STATE {\em Post-processing after the iteration}:\
Use
\cite[Algorithms 1 and 2]{ncjs22}
to implement the filtering procedure
${\bf w}= {\bf R} {\bf H} {\bf z}_1={\bf W}_{\rm mse}^0 {\bf y}$ at the vertex $i$, where the input
is  $z_1^{(L_1)}(i)$ and the output
denoted by $w^{(L_1)}(i)$, is
the approximate value of the output data of the above filtering procedure.

\STATE {\bf Part II}: \ Implementation of the regularization procedure \eqref{stochasticrecovery.eqb} at the vertex $i$

\STATE {\em Pre-processing}: \ Rescaling $z_2^{(L_1)}(i)= p(i)^{1/2} w^{(L_1)}(i)$,
the approximate value of the output data of the rescaling procedure ${\bf z}_2= {\bf P}^{1/2} {\bf w}$.

\STATE {\em  Iteration}: \ Start from ${\bf w}_0(i)=z_2^{(L_1)}(i)$, and
use \cite[Algorithms 1 and 2]{ncjs22} and rescaling ${\bf P}^{-1/2}$ to
implement the  procedure \eqref{PKinverse.eq2} for $0\le m\le L_2$, with
the output, denoted by $z_3^{(L_1, L_2)}(i)$,
being the approximation value of the output data of the inverse
filtering procedure ${\bf z}_3=\big({\bf I}+{\bf P}^{-1/2} {\bf K}{\bf P}^{-1/2})^{-1} {\bf z}_2$
at the vertex $i$

\STATE {\em Post-processing}:\ ${ x}_{\rm mse}^{(L_1, L_2)}(i)=p(i)^{-1/2}  z_3^{(L_1, L_2)}(i)$.

\STATE {\em Output}: $ x_{\rm mse}(i)\approx  { x}_{\rm mse}^{(L_1, L_2)}(i)$, the approximate value of the output data of
 the Wiener filtering procedure $ {\bf x}_{\rm mse}= ({\bf P}+{\bf K})^{-1} {\bf P}{\bf w}={\bf W}_{\rm mse} {\bf y}$
 at the vertex $i$.

\end{algorithmic}  %\vspace{-.03in}
\end{algorithm}

\begin{remark} {\rm
We remark that  for the case that  the probability measure $P$ is uniform
 \cite{perraudin17},
 ${\bf I}+{\bf P}^{-1/2} {\bf K}{\bf P}^{-1/2}={\bf I}+N {\bf K}$ is a polynomial filter of ${\bf S}_1, \ldots, {\bf S}_d$
 if ${\bf K}=k({\bf S}_1, \ldots, {\bf S}_d)$ is, and hence JPA$(\alpha, \beta)$ and  CIPA algorithms
 %\st{ the Jacobi polynomial approximation algorithm} \eqref{jacobiapproximation.def}
  proposed in Section \ref{Jacobiapproximation.section}
 can be applied to the inverse filtering procedure ${\bf z}_3=\big({\bf I}+{\bf P}^{-1/2} {\bf K}{\bf P}^{-1/2})^{-1} {\bf z}_2$
 if $ 1+ N k({\bf t})>0 \ {\rm for \ all} \ {\bf t}\in [{\pmb \mu}, {\pmb \nu}]$.
  }\end{remark}

We finish this section with optimal {\bf  unbiased}  Wiener filters for the scenario that the input signals
${\bf x}$  are  wide-stationary, i.e., ${\bf x}$ satisfies \eqref{wiener2.eq4},
 %{\color{red} ${\bf 1}$ has been defined in the notation section} \st{ is the column vector with all entries taking value one},
%i.e., \eqref{wiener2.eq4} is satisfied,
the filtering procedure satisfies \eqref{wiener2.eqb} and
\begin{equation}\label{wbs.reuqirement1} {\bf H}{\bf 1}=\tau {\bf 1}\end{equation}
for some  $\tau\ne 0$, the output ${\bf y}$ in \eqref{wiener2.eq2} are corrupted by some noise ${\pmb \epsilon}$ satisfying
\eqref{wiener2.eq3}, and the covariance matrix ${\bf G}$ of the noise and the regularization matrix ${\bf K}$ satisfy
\begin{equation}\label{KG.condition}
{\bf G}{\bf 1}={\bf K}{\bf 1}={\bf 0}.
\end{equation}
In the above setting, the random variable $\tilde {\bf x}={\bf x}-{\mathbb E}{\bf x}={\bf x}- c {\bf 1}$ satisfies
\begin{equation}\label{wiener2.eq4b}
{\mathbb E}\tilde {\bf x}={\bf 0},  {\mathbb E} {\tilde {\bf x}}{\pmb \epsilon}^T={\bf 0}\ {\rm and}
 \  {\mathbb E} \tilde {\bf x} \tilde {\bf x}^T= \widetilde{\bf R}=\tilde r({\mathbf S}_1, \ldots, {\mathbf S}_d).
\end{equation}
%{\color{red} where $\widetilde{\bf R}= \tilde r({\mathbf S}_1, \ldots, {\mathbf S}_d)$ is the covariance matrix of $\bf x$.}
For any unbiased reconstruction filter ${\bf W}$, we have
$${\bf W}{\bf H} {\bf 1}={\bf 1}.$$
This together with \eqref{KG.condition} implies that
\begin{eqnarray*}% \label{wiener2.eq2b}
{\bf W} {\bf y}-{\bf x}  & \hskip-0.08in = & \hskip-0.08in c ({\bf W}{\bf H} {\bf 1}-{\bf 1})+ ({\bf W}{\bf H}-{\bf I})\tilde {\bf x}+{\bf W}{\pmb \epsilon}\nonumber\\
 & \hskip-0.08in = & \hskip-0.08in  ({\bf W}{\bf H}-{\bf I})\tilde {\bf x}+{\bf W}{\pmb \epsilon}
\end{eqnarray*}
and
\begin{eqnarray*}% \label{wiener2.eq2c}
{\bf y}^T {\bf W}^T {\bf K} {\bf W}{\bf y}  & \hskip-0.08in = & \hskip-0.08in
({\bf H}\tilde {\bf x}+{\pmb \epsilon})^T{\bf W}^T{\bf K} {\bf W} ({\bf H}\tilde {\bf x}+{\pmb \epsilon})+{\bf 1}^T {\bf K}{\bf 1}\nonumber\\
& \hskip-0.08in & \hskip-0.08in
+{\bf 1}^T {\bf K} {\bf W}({\bf H}\tilde {\bf x}+{\pmb \epsilon}) +
({\bf H}\tilde {\bf x}+{\pmb \epsilon})^T {\bf W}^ T {\bf K}{\bf 1}   \nonumber\\
 & \hskip-0.08in = & \hskip-0.08in
({\bf H}\tilde {\bf x}+{\pmb \epsilon})^T{\bf W}^T{\bf K} {\bf W} ({\bf H}\tilde {\bf x}+{\pmb \epsilon}).
\end{eqnarray*}
Therefore following the argument used in the proof of Theorem
\ref{wienerfiltermse.thm} with the signal ${\bf x}$ and polynomial $r$ replaced by $\tilde{\bf x}$ and $\tilde r$ respectively,
and applying \eqref{stochasticestimator.unbiased}, \eqref{wbs.reuqirement1} and \eqref{wiener2.eq4b}, we can show that
the stochastic Wiener filter $\widetilde{\bf W}_{\rm mse}$ in \eqref{stationarywienerfiltermse.eq2}
 is
 an optimal unbiased filter to reconstruct wide-band stationary signals.

\begin{thm}\label{widebandwienerfilter.thm}
Let the input signal ${\bf x}$, the noisy output signal ${\bf y}$ and the additive noise ${\pmb \epsilon}$  be
  in \eqref{wiener2.eq4}, \eqref{wiener2.eq2}, \eqref{wiener2.eq3},  the covariance matrix ${\bf G}$ of the noise and the regularization matrix ${\bf K}$ satisfy
\eqref{KG.condition}, and let the filtering procedure associated with the filter ${\bf H}$  satisfy \eqref{wiener2.eqb} and \eqref{wbs.reuqirement1}.
Assume that
${\bf H}\widetilde {\bf R}{\bf H}^T+{\bf G}$  and ${\bf P}+{\bf K}$ are strictly positive definite. Then
\begin{equation} \label{stationarywienerfiltermse.eq1}
F_{{\rm mse}, P, {\bf K}}({\bf W})\ge  F_{{\rm mse}, P, {\bf K}}(\widetilde {\bf W}_{\rm mse})
\end{equation}
hold for all unbiased reconstructing filters ${\bf W}$, where $F_{{\rm mse}, P, {\bf K}}({\bf W})$ is the stochastic  mean squared error   in  \eqref{mse.objectivefunction} and
\begin{equation}  \label{stationarywienerfiltermse.eq2}
\widetilde{\bf W}_{\rm mse}=({\bf P}+{\bf K})^{-1}{\bf P}\widetilde {\bf R}{\bf H}^T ({\bf H}\widetilde {\bf R}{\bf H}^T+{\bf G})^{-1}.
\end{equation}
Moreover,
$\tilde {\bf x}_{\rm mse}=\widetilde {\bf W}_{\rm mse}{\bf y}$
is an unbiased estimator to the wide-band stationary signal ${\bf x}$.
\end{thm}

Following the distributed algorithm used to implement
the stochastic Wiener filtering procedure,
the unbiased estimation
$\tilde {\bf x}_{\rm mse}=\widetilde {\bf W}_{\rm mse}{\bf y}$
 can be implemented at the vertex level with one-hop communication %\st{data exchanging with adjacent vertices only}
 when
  \begin{equation*}%\label{stochasticwienerfilter.widebandsignalcondition}
   h^2({\bf t}) \tilde r({\bf t})+g({\bf t})>0 \ {\rm for \ all} \ {\bf t}\in [{\pmb \mu}, {\pmb \nu}].\end{equation*}
Numerical demonstrations to denoise wide-band stationary signals are presented in Section  \ref{denoisingwideband.demo}. %{simulation.section}

\section{Wiener filters for  deterministic graph signals}
\label{worst-casewienerfilter.section}

Let ${\mathbf S}_1, \ldots, {\mathbf S}_d$ be real  commutative  symmetric graph shifts on a simple graph ${\mathcal G}=(V, E)$
and their joint spectrum be contained in some cube
$[{\pmb \mu}, {\pmb \nu}]$, i.e., \eqref{jointspectralcubic.def} holds.
In this section, we consider the scenario that
the filtering procedure \eqref{filtering.def}
has the  filter  ${\bf H}$ given in \eqref{wiener2.eqb}, % a
its inputs
${\bf x}=(x(i))_{i\in V}$ are
 deterministic signals with their energy bounded by some $\delta_0>0$,
\begin{equation} \label{wiener1.eqa1}
\|{\bf x}\|_2\le \delta_0, \end{equation}
 and
  its outputs
 \begin{equation}\label{wiener1.eqc}
{\bf y}={\bf H} {\bf x}+{\pmb \epsilon}\end{equation}
 are corrupted by some random noise ${\pmb \epsilon}$ which has mean zero and covariance matrix  ${\bf G}={\rm cov}({\pmb \epsilon})$ being
  a polynomial of graph shifts ${\bf S}_1, \ldots, {\mathbf S}_d$,
\begin{equation} \label{wiener1.eqd} {\mathbb E}{\pmb \epsilon}={\bf 0}\ \ {\rm and}\  {\bf G}=g({\bf S}_1, \ldots, {\bf S}_d)
\end{equation}
for some multivariate polynomial $g$.
For the above  setting of the filtering procedure, we
 introduce
 the {\em  worst-case mean squared error}
of  a reconstruction filter ${\bf W}$
by
\begin{equation}\label{wcms.objectivefunction}
F_{{\rm wmse}, P}({\bf W})= \sum_{i\in V} p(i) \max_{\|{\bf x}\|_2\le \delta_0}
{\mathbb E}  | ({\bf W} {\bf y})(i)-{ x}(i)|^2,
\end{equation}
where  $P=(p(i))_{i\in V}$ is  a  probability measure  on the graph ${\mathcal G}$ \cite{bi2009, eldar2006}.
 In this section, we
discuss the optimal
 reconstruction filter  ${\bf W}_{\rm wmse}$ with respect to
 the worst-case mean squared error  $F_{{\rm wmse}, P}$ in \eqref{wcms.objectivefunction},
   and we
 propose a distributed algorithm to implement the worst-case Wiener filtering procedure
 % the signal recovery procedure  $\tilde {\bf x}={\bf W}_{\rm wmse}{\bf y}$
 at the vertex
level with one-hop communication. % with data exchanging with adjacent vertices only.

First,  we provide a {\bf universal} solution to the  minimization problem
 \begin{equation} \label{wcms.minimization}
\min_{\bf W} F_{{\rm wmse}, P}({\bf W}),
\end{equation}
which is independent of the probability measure $P$, see Appendix \ref{wienerfilterworsecase.thm.pfappendix} for the proof.

\begin{thm} \label{wienerfilterworsecase.thm}
Let  the filter   ${\bf H}$, the input ${\bf x}$, the noisy output  ${\bf y}$,
the noise $\pmb \epsilon$,  and  the worst-case mean squared error $F_{{\rm wmse}, P}$
 be as in \eqref{wiener2.eqb}, \eqref{wiener1.eqa1}, \eqref{wiener1.eqc}, \eqref{wiener1.eqd} and \eqref{wcms.objectivefunction}
  respectively. %\eqref{wiener1.eq},
% \eqref{wiener1.eqb}
%\eqref{wiener1.eqa1},  \eqref{wiener1.eqc}  and \eqref{wiener1.eqd},
Assume that   $\delta_0^2 {\bf H}{\bf H}^T+{\bf G}$ is strictly positive definite. Then
\begin{eqnarray} \label{wienerfilterworsecase.eq1}
F_{{\rm wmse}, P}({\bf W}) & \hskip-0.08in \ge & \hskip-0.08in   F_{{\rm wmse}, P}({\bf W}_{\rm wmse})\nonumber\\
& \hskip-0.08in = & \hskip-0.08in
  \delta_0^2- \delta_0^4 {\rm tr} ( (\delta_0^2 {\bf H} {\bf H}^T +{\bf G})^{-1} {\bf H} {\bf P} {\bf H}^T \big)\qquad
\end{eqnarray}
hold for all reconstructing filters ${\bf W}$, where
 ${\bf P}$ is the diagonal matrix with diagonal entries $p(i), i\in V$, and
\begin{equation}\label{wienerfilterworsecase.eq2}
{\bf W}_{{\rm wmse}}= \delta_0^2 {\bf H}^T \big(\delta_0^2 {\bf H}{\bf H}^T+{\bf G})^{-1}.
\end{equation}
 Moreover,
the reconstruction filter
${\bf W}_{\rm wmse}$
is the unique solution of the minimization problem
\eqref{wcms.minimization} if ${\bf P}$ is invertible, i.e.,
the probability $p(i)$ at every vertex  $i\in V$ is positive. % is po, i.e., $p(i)>0$ for all $i\in V$.
\end{thm}

We call the optimal reconstruction error ${\bf W}_{\rm wmse}$ in \eqref{wienerfilterworsecase.eq2}
as the {\em worst-case Wiener filter}.
Denote the order of the graph ${\mathcal G}$ by $N$.
For the case that the probability measure $P$ is the uniform probability measure $P_U$,
we can simplify the estimate \eqref{wienerfilterworsecase.eq1}
%of
%the worst-case mean squared error
%  for the worst-case Wiener filter ${\bf W}_{\rm wmse}$
as follows:
\begin{equation}\label{wienerfilterworsecase.error2}
F_{{\rm wmse}, P_U}({\bf W}_{\rm wmse})=\frac{\delta_0^2}{N} {\rm tr}  \big( (\delta_0^2 {\bf H} {\bf H}^T +{\bf G})^{-1} {\bf G}\big),
\end{equation}
c.f. \eqref{fmsepu}.
If the random noises ${\pmb \epsilon}$ are further assumed to be i.i.d  and have mean zero and variance $\sigma$,  we can use
singular values $\mu_i({\bf H}), 1\le i\le N$, of the filter ${\bf H}$ to
estimate the worst-case mean squared error  for the worst-case Wiener filter ${\bf W}_{\rm wmse}$,
\begin{equation}\label{wienerfilterworsecase.error3}
F_{{\rm wmse}, P_U}({\bf W}_{\rm wmse})=\frac{\delta_0^2 \sigma^2}{N}  \sum_{i=1}^N \frac{1} {\delta_0^2 \mu_i({\bf H})^2+\sigma^2}.
\end{equation}

Denote the reconstructed signal via the worst-case Wiener filter ${\bf W}_{\rm wmse}$ by
\begin{equation} {\bf x}_{\rm wmse}={\bf W}_{\rm wmse} {\bf y},\end{equation}
where ${\bf y}$ is given in \eqref{wiener1.eqc}.
By  \eqref{wienerfilterworsecase.eq2}, the reconstructed signal ${\bf x}_{\rm wmse}$ can be obtained by the combination
of an inverse filtering procedure
\begin{subequations} \label{worstcaserecovery.eq}
\begin{equation} \label{worstcaserecovery.eqb}
{\bf z} = \big(\delta_0^2 {\bf H} {\bf H}^T +{\bf G}\big)^{-1} {\bf y}
\end{equation}
and a filtering procedure
\begin{equation} \label{worstcaserecovery.eqa}  {\bf x}_{\rm wmse}= \delta_0^2 {\bf H}^T {\bf z},
\end{equation}
\end{subequations}
where the noisy observation ${\bf y}$ is  the input and
$\delta_0^2 {\bf H} {\bf H}^T +{\bf G}$ is a polynomial filter.
As the graph shifts ${\mathbf S}_1, \ldots, {\mathbf S}_d$ are symmetric and commutative,  $\bf H$ is a polynomial graph filter in   \eqref{wiener2.eqb} and  \eqref{wiener1.eqd} holds, we have that
 ${\bf H}^T={\bf H}=h({\mathbf S}_1, \ldots, {\mathbf S}_d)$
 and
 $\delta_0^2 {\bf H} {\bf H}^T +{\bf G}= \delta_0^2 {\bf H}^2 +{\bf G}= (\delta_0^2 h^2+g)({\mathbf S}_1, \ldots, {\bf S}_d)$
 are polynomial filters of ${\bf S}_1, \ldots, {\bf S}_d$.
Therefore using \cite[Algorithms 1 and  2]{ncjs22}, the filtering procedure
\eqref{worstcaserecovery.eqa} can be implemented at the vertex level  with one-hop communication.
 By Theorem \ref{exponentialconvergence.thm},
 the  polynomial approximation algorithm \eqref{jacobiapproximation.def} proposed in the last section
 can be applied to the inverse filtering procedure \eqref{worstcaserecovery.eqb}
  if the following requirement is met,
 \begin{equation*}\label{worsecasewienerfilter.filtercondition}
  \delta_0^2 h^2({\bf t})+g({\bf t})>0 \ {\rm for \ all} \ {\bf t}\in [{\pmb \mu}, {\pmb \nu}].\end{equation*}
Hence %under the additional assumption
% \eqref{worsecasewienerfilter.filtercondition}
% on the polynomials $h$ and $g$,
 the worst-case Wiener filtering procedure
  \eqref{worstcaserecovery.eq}
 can be implemented at the vertex level with one-hop communication, see  Algorithm \ref{Wiener1.algorithm}
 for the implementation at a vertex. % The numerical demonstration  for signal reconstruction via the worst-case Wiener filter ${\bf W}_{\rm wmse}$

      \begin{algorithm}[t]
\caption{Polynomial approximation algorithm  to implement the  worst-case Wiener filtering procedure $ {\bf x}_{\rm wmse}={\bf W}_{\rm wmse} {\bf y}$
  at a vertex $i\in V$. }
\label{Wiener1.algorithm}
\begin{algorithmic}  %[1]

\STATE {\em Inputs}: Polynomial coefficients of polynomial filters ${\bf H}, {\bf G}$ and ${\bf G}_M$ (either Jacobi polynomial approximation filter  ${\bf G}_M^{(\alpha, \beta)}$ or Chebyshev interpolation approximation filter ${\bf C}_M$),  entries $S_k(i,j), j\in {\mathcal N}_i$ in the $i$-th row of the shifts ${\bf S}_k, 1\le k\le d$,
the  value $y(i)$  of the input signal ${\bf y}=(y(i))_{i\in V}$ at the vertex $i$, and number $L$ of iteration.

%\STATE {\bf Operation}: Evaluate $m_k=\mu(B(k, r))$, compute ${\bf F}_k= {\bf H}_{0,k}^T{\bf H}_{0,k}+ {\bf H}_{1,k}^T{\bf H}_{1,k}$,
% find its inverse  $({\bf F}_k)^{-1}$, and then compute $ {\bf G}^L_{l; k}:=({\bf F}_k)^{-1} {\bf H}_{l,k}^T, l=0, 1$.

\STATE {\em Pre-iteration}: Find the polynomial coefficients of polynomial filter $\delta_0^2 {\bf H}^2+{\bf G}$.

\STATE {\em Initialization}:   $n=0$ and zero initial  $x^{(0)}(i)=0$.

\STATE{\em Iteration}:  Use
\cite[Algorithms 1 and 2]{ncjs22}
to implement the filtering procedures
${\bf e}^{(m)}=(\delta_0^2 {\bf H}^2+{\bf G}) {\bf x}^{(m-1)}-{\bf y}$ and
${\bf x}^{(m)}= {\bf x}^{(m-1)}-{\bf G}_M {\bf e}^{(m)}$ at the vertex $i$,
with the output of the $L$-th iteration denoted by
$ x^{(L)}(i)$.

\STATE {\em Post-iteration}: Use
\cite[Algorithms 1 and  2]{ncjs22}
to implement the filtering procedure
$ {\bf x}_{\rm wmse}=\delta_0^2 {\bf H} {\bf x}^{(L)}$ at the vertex $i$,
with the output  denoted by $ x_{\rm wmse}^{(L)}(i)$.

\STATE {\em Output}: $ x_{\rm wmse}(i)\approx  x^{(L)}_{\rm wmse}(i)$, the approximate value of the output data of
 the Wiener filtering procedure $ {\bf x}_{\rm wmse}={\bf W}_{\rm wmse} {\bf y}$
 at the vertex $i$.
\end{algorithmic}  %\vspace{-.03in}
\end{algorithm}

 For a  probability measure  $P=(p(i))_{i\in V}$ on the graph ${\mathcal G}$ and
 a reconstruction filter ${\bf W}$,
\begin{eqnarray}\label{wcms.objectivefunction2}
\widetilde F_{{\rm wmse}, P}({\bf W})= \max_{\|{\bf x}\|_2\le \delta_0} \sum_{i\in V} p(i)
{\mathbb E}  | ({\bf W} {\bf y})(i)-{\bf x}(i)|^2
\end{eqnarray}
is another natural   %{\color{red} conventional?}
 worst-case mean squared error measurement, c.f. \eqref{wcms.objectivefunction}.
%Let ${\bf P}$ be the diagonal matrix with diagonal entries $p(i), i\in V$.
By \eqref{wiener1.eqc} and \eqref{wiener1.eqd}, we obtain
\begin{eqnarray*}%\label{wcms.objectivefunction2.bound1}
  & \hskip-0.08in & \hskip-0.08in   \widetilde F_{{\rm wmse}, P}({\bf W}) \nonumber\\
  & \hskip-0.08in =& \hskip-0.08in
\sup_{\|{\bf x}\|_2\le \delta_0}  {\bf x}^T ({\bf H}^T {\bf W}^T-{\bf I}) {\bf P} ({\bf W}{\bf H}-{\bf I}) {\bf x}\nonumber\\
& & + {\rm tr} \big({\bf P} {\bf W} ({\mathbb E}({\pmb \epsilon} {\pmb \epsilon}^T){\bf W}^T\big)\nonumber\\
& \hskip-0.08in = &  \hskip-0.08in \delta_0^2 \lambda_{\max} \left(({\bf H}^T {\bf W}^T-{\bf I}) {\bf P} ({\bf W}{\bf H}-{\bf I})\right) + {\rm tr} ({\bf P} {\bf W} {\bf G}{\bf W}^T)\nonumber\\
&  \hskip-0.08in \le  &  \hskip-0.08in \delta_0^2 {\rm tr} \left(({\bf H}^T {\bf W}^T-{\bf I}) {\bf P} ({\bf W}{\bf H}-{\bf I})\right) + {\rm tr} ({\bf P} {\bf W} {\bf G}{\bf W}^T)\nonumber\\
&  \hskip-0.08in = &  \hskip-0.08in  {\rm tr} \left( {\bf P} \big(\delta_0^2 ({\bf W}{\bf H}-{\bf I})({\bf H}^T {\bf W}^T-{\bf I})+ {\bf W} {\bf G}{\bf W}^T\big)\right)\nonumber\\
& \hskip-0.08in = &  \hskip-0.08in F_{{\rm wmse}, P}({\bf W}),
\end{eqnarray*}
where the inequality holds as the matrix $({\bf H}^T {\bf W}^T-{\bf I}) {\bf P} ({\bf W}{\bf H}-{\bf I})$ is positive semidefinite.
Similarly, we have the following lower bound estimate,
\begin{eqnarray*}% \label{wcms.objectivefunction2.bound2}
 \hskip-0.18in \widetilde F_{{\rm wmse}, P}({\bf W})
&  \hskip-0.08in \ge  &  \hskip-0.08in \frac{\delta_0^2}{N} {\rm tr} \big(({\bf H}^T {\bf W}^T-{\bf I}) {\bf P} ({\bf W}{\bf H}-{\bf I})\big) \nonumber\\
 \hskip-0.18in  &  \hskip-0.08in   &  \hskip-0.08in  + {\rm tr} ({\bf P} {\bf W} {\bf G}{\bf W}^T)\ge \frac{F_{{\rm wmse}, P}({\bf W})}{N}.
\end{eqnarray*}
For the case that the probability measure is  uniform  and  the random noise vector $\pmb \epsilon$ is i.i.d. with  mean zero and variance $\sigma^2$,
we get
\begin{eqnarray*}
 %& \hskip-0.08in & \hskip-0.08in
  \widetilde F_{{\rm wmse}, P_U}({\bf W}_{\rm wmse})  %\nonumber\\
% & \hskip-0.08in = & \hskip-0.08in
%\delta_0^2 \lambda_{\max} \big(({\bf H}^T {\bf W}_{\rm wor}^T-{\bf I}) {\bf P} ({\bf W}_{\rm wor}{\bf H}-{\bf I})\big)\nonumber\\
%& & + {\rm tr} ({\bf P} {\bf W}_{\rm wor} {\bf G}{\bf W}_{\rm wor}^T)\nonumber\\
%& \hskip-0.08in = & \hskip-0.08in \frac{\delta_0^2}{N} \Big(\lambda_{\max}\big( \delta_0^2 {\bf H}^T (\delta_0^2 {\bf H}{\bf H}^T+\sigma^2 {\bf I})^{-1} {\bf H}-{\bf I}\big)\Big)^2\nonumber\\
%& & + \frac{\delta_0^4\sigma^2}{N}{\rm tr} ({\bf H}^T (\delta_0^2 {\bf H}{\bf H}^T+\sigma^2 {\bf I})^{-2}{\bf H}\big)\nonumber\\
& \hskip-0.08in = & \hskip-0.08in \frac{\delta_0^2 \sigma^2}{N} \max_{1\le i\le N} \frac{\sigma^2}{(\delta_0^2 \mu_i({\bf H})^2+\sigma^2)^2}\nonumber\\
& &   \hskip-0.08in +
\frac{\delta_0^2\sigma^2}{N}  \sum_{i=1}^N \frac{\delta_0^2 \mu_i({\bf H})^2}{ (\delta_0^2 \mu_i({\bf H})^2+\sigma^2)^2},\qquad \quad
\end{eqnarray*}
where $\mu_i({\bf H}), 1\le i\le N$, are singular values of the filter ${\bf H}$, cf. \eqref{wienerfilterworsecase.error3}
for the estimate for $F_{{\rm wmse}, P_U}({\bf W}_{\rm wmse})$.

\section{Simulations}
\label{simulation.section}

	Let $N\ge 1$ and we say that
 $a=b\ {\rm mod }\ N$ if $(a-b)/N$ is an integer.
 The  {\em circulant graph} ${\mathcal C}(N, Q)$  generated by  $Q=\{q_1, \ldots, q_L\}$
is a simple graph with  the vertex set   $V_N=\{0, 1, \ldots, N-1\}$  and the edge set
%\begin{equation}\label{circulant.edgedef}
$E_N(Q)=\{(i, i\pm q\ {\rm mod}\ N),\  i\in V_N, q\in Q\}$,  % \end{equation}
where   $q_l, 1\le l\le L$, are integers contained in $[1, N/2)$
  \cite{ncjs22, ekambaram13}-\cite{dragotti19}.
%  \cite{ncjs22, ekambaram13, vnekambaram13, dragotti19a, dragotti19}.
%   Circulant graphs are widely used in image processing
%\cite{ekambaram13, vnekambaram13, dragotti19a, dragotti19}
 In Section \ref{circulantgraph.demo},   we demonstrate the theoretical result in Theorem  \ref{exponentialconvergence.thm} on
 the exponential convergence of  the  Jacobi polynomial approximation algorithm (JPA($\alpha, \beta$))
and  Chebyshev interpolation polynomial algorithm (CIPA)
on the implementation of  inverse filtering procedures on circulant graphs. Our numerical results  show that
  the CIPA  %Chebyshev interpolation polynomial approximation algorithms CIPA and
%Jacobi polynomial approximation algorithms
  and JPA($\alpha, \beta$) with appropriate selection of parameters $\alpha$ and $\beta$
have superior performance to implement the inverse procedure than
the Chebyshev polynomial approximation algorithm  in \cite{ncjs22} and
the gradient descent method  in \cite{Shi15} do.

Let   ${\mathcal G}_{N}=(V_{N}, E_{N}), N\ge 2$, be  random geometric graphs with vertices randomly deployed on
 $[0, 1]^2$ and
  an undirected edge between two vertices if their physical
distance is not larger than
$\sqrt{2/N}$  \cite{ncjs22, jiang19, Nathanael2014}.
In Sections \ref{randomsignal.demo} and \ref{denoisingwideband.demo}, we consider  denoising (wide-band) stationary signals
via the Wiener procedures with/without regularization taken into account, % \eqref{stochasticestimator}, and
 and we compare the performance of denoising via %the Wiener procedure \eqref{stochasticrecovery.eqa}
% without regularization taken into the account and
 the Tikhonov
regularization method \eqref{tik.def}.
It is observed that the Wiener filtering procedures
with/without regularization taken into account
 have better performance on denoising (wide-band) stationary signals than the conventional Tikhonov regularization approach does.
%, especially signals with certain regularity.

\subsection{Polynomial approximation algorithms on circulant graphs}
\label{circulantgraph.demo}

 In  simulations of this subsection, we take  circulant graphs  ${\mathcal C}(N,Q_0)$, polynomial filters  ${\bf H}_1$, input signals ${\bf x}$ of
the filtering procedure ${\bf x}\longmapsto {\bf H}_1{\bf x}$, and  input signals ${\bf y}$ of the inverse filtering procedure
 ${\bf y}\longmapsto  {\bf H}_1^{-1}{\bf y}$  as in \cite{ncjs22}, that is,
  the circulant graphs ${\mathcal C}(N,Q_0)$  are
 generated by $Q_0 = \{1, 2, 5\}$, % (see \cite[Fig. 2]{ncjs22} for the circulant graph $C(N,Q_0)$ with $N=50$),
  ${\bf H}_1=h_1( {\bf L}_{C(N,Q_0)}^{\rm sym})$
 is  a polynomial filter of the symmetric normalized Laplacian ${\bf L}_{C(N,Q_0)}^{\rm sym}$ on the circulant graph   ${\mathcal C}(N,Q_0)$
 with $ h_1(t)=(9/4-t)(3+t)$  given in \eqref{h1.def},
   the input signal ${\bf x}$ has i.i.d. entries randomly selected in $[-1, 1]$,
    and  the  input signal  ${\bf y}={\bf H}_1{\bf x}$
  of the inverse filtering procedure is the output of the filtering procedure. Shown in Table
\ref{CirculantGraphICPA.Table} are
averages of the relative  iteration error
$${\rm E}(m)=\frac{ \|{\bf x}^{(m)}-{\bf x}\|_2}{\|{\bf x}\|_2},\  m\ge 1,$$
over 1000 trials to implement the inverse filtering procedure ${\bf y}\longmapsto {\bf H}_1^{-1} {\bf y}$ via
 the  JPA($\alpha, \beta$)
and   CIPA with zero initial  ${\bf x}^{(0)}={\bf 0}$,
where ${\bf x}^{(m)}, m\ge 1$, are the output of the polynomial approximation algorithm
\eqref{jacobiapproximation.def} at $m$-th iteration and   $M$ is  the degree of polynomials in the Jacobi (Chebyshev interpolation) polynomial approximation.
% and Chebyshev interpolation polynomial approximation.  %  $0\le M\le 4$.

The   JPA($\alpha, \beta$) with $\alpha=\beta=-1/2$ is the Chebyshev
polynomial approximation algorithm,  ICPA for abbreviation,  introduced in \cite{ncjs22}
and the relative iteration error presented in Table \ref{CirculantGraphICPA.Table}
for the  JPA($-1/2, -1/2$) is copied from \cite[Table 1]{ncjs22}.
We observe %from Table \ref{CirculantGraphICPA.Table}
that
%Chebyshev interpolation polynomial approximation algorithms
CIPA and
%Jacobi polynomial approximation algorithms
 JPA($\alpha, \beta$) with appropriate selection of parameters $\alpha$ and $\beta$
 have better performance on the implementation of inverse filtering procedure
than the
% Chebyshev polynomial approximation algorithm
 ICPA in \cite{ncjs22} does,
and
they %the proposed Chebyshev interpolation polynomial approximation algorithms
%CIPA and
%Jacobi polynomial approximation algorithms  JPA($\alpha, \beta$)
have much better performance if we select  approximation polynomials with higher order $M$.   % the approximation filter is selected.

As the filter ${\bf H}_1$ is a positive definite matrix,  the inverse filtering procedure ${\bf y}\longmapsto {\bf H}_1^{-1}{\bf y}$
can also be implemented by the  gradient descent method  with optimal step size \eqref{gd0.def},  GD0 for abbreviation
\cite{Shi15}.
%$${\bf x}^{(m)} = {\bf x}^{(m-1)}- \gamma_{\rm opt}  ({\bf H}_1 {\bf x}^{(m-1)}- {\bf y}),  m\ge 1, $$
%where $\gamma_{\rm opt}=2/(\lambda_{\min} + \lambda_{\max})$ and $\lambda_{\min}$ (resp. $\lambda_{\max}$)
% are the minimal (resp. maximal) eigenvalues of the matrix ${\bf H}_1$
%The above gradient descent method with GD0 for abbreviation can be considered as a polynomial approximation algorithm \eqref{Approximationalgorithm} with the approximation %filter ${\bf G}=\gamma_{\rm opt} {\bf I}$ \cite{jiang19, Leus17, Waheed18, Shi15, sihengTV15, Shuman18, isufi19}.
Shown in the sixth row of  Table \ref{CirculantGraphICPA.Table}, which is copied from \cite[Table 1]{ncjs22},
is the relative  iteration error to implement the inverse filtering ${\bf y}\longmapsto {\bf H}_1^{-1}{\bf y}$.
It indicates that  the CIPA  %Chebyshev interpolation polynomial approximation algorithms CIPA
and
%Jacobi polynomial approximation algorithms
 JPA($\alpha, \beta$) with appropriate selection of parameters $\alpha$ and $\beta$
have superior performance to implement the inverse procedure than
the  gradient descent method  does. %\st{ even as observed in} \cite[Table 1]{ncjs22} \st{that
% it has  better performance than the ICPA does, see the relative iteration error listed in the first row of
%  Table} \ref{CirculantGraphICPA.Table}.  %superierhas better performance than the

\begin{table}[t]
		\centering
		\caption{ 		Average   relative iteration errors  $E(m)$
to implement the inverse filtering  ${\bf y}\longmapsto {\bf H}_1^{-1} {\bf y}$ on the circulant graph ${\mathcal C}(1000, Q_0)$ via polynomial approximation algorithms
and the gradient descent method with zero initial. }
		\label{CirculantGraphICPA.Table}			
       \begin{tabular} {|c|c|c|c|c|c|c|c|}	
   			\hline
   \hline
			\backslashbox{Alg.} %orithm} %$(\alpha, \beta, M)$}
%{ME}
{Iter. $m$} %ation}
& 1 & 2 & 3 & 4 & 5 \\ % & 6 &7&8&9 &10 &12&16&20 \\
			\hline
			\hline
     \multicolumn{6}{c}{\multirow{1}{*}{$M=0$}}\\
     \hline
           \hline
% {$\eta$=3/4, ISNR=  3.3755} %\vline
%\hline
JPA(-${1}/{2}$, -${1}/{2}$)	&
0.5686  &  0.4318  &  0.3752  &  0.3521 &   0.3441\\
%  &  0.3422   & 0.3449&  0.3500  &  0.3565  &  0.3643 &   0.3828  &  0.4296 &   0.4890\\
\hline
JPA(${1}/{2}$, ${1}/{2}$)	&	
0.3007 	&   0.1307  	&  0.0677	&    0.0379  	&  0.0219  \\
%	&  0.0129  	&  0.0076		&  0.0045   &  0.0027   &  0.0016  &   0.0006  &   0.0001 &    0.0000\\
\hline
JPA(${1}/{2}$,-${1}/{2}$)&
0.2298 &   0.0955&    0.0452 &   0.0223   & 0.0113 \\
%  &  0.0057  &  0.0029& 0.0015 &  0.0008 &   0.0004  &  0.0001  &  0.0000  &  0.0000\\
\hline
JPA(0,-${1}/{2}$)&
   0.2296  &  0.0833 &   0.0337 &   0.0141  &  0.0060 \\
   %&   0.0026    &0.0011&   0.0005  &  0.0002  &  0.0001  &  0.0000 &   0.0000  &  0.0000 \\
\hline	
CIPA  &
 0.2189  &  0.0822 &  0.0347  &  0.0154  &  0.0070 \\
 \hline
 % &  0.0033   & 0.0015& 0.0007  &  0.0003 &   0.0002  &  0.0000 &   0.0000 &   0.0000\\
% ARMA & 0.3259 & 0.2583 & 0.1423 & 0.1098 & 0.0718\\
% \hline
GD0 & 0.2350 & 0.0856 & 0.0349 & 0.0147 & 0.0063\\
 \hline
 			\hline
     \multicolumn{6}{c}{\multirow{1}{*}{$M=1$}}\\ \hline
           \hline
 JPA(-${1}/{2}$, -${1}/{2}$)	&
0.4494   & 0.2191 &   0.1103  &  0.0566   & 0.0295 \\
%&   0.0155  &  0.0082& 0.0044 &   0.0024  &  0.0013 &   0.0004   & 0.0000  &  0.0000\\
\hline
 JPA(${1}/{2}$, ${1}/{2}$)
	&  0.2056  &  0.0769 &   0.0390 &   0.0213 &   0.0119\\
% &    0.0067 &   0.0038&   0.0022   & 0.0012  &  0.0007   & 0.0002 &   0.0000  & 0.0000\\
\hline
JPA(${1}/{2}$, -${1}/{2}$)
& 0.1624   & 0.0297 &   0.0056 &   0.0011 &   0.0002 \\
% &  0.0000  &  0.0000&  0.0000  &  0.0000 &   0.0000  &  0.0000   & 0.0000   & 0.0000
\hline
JPA(0, -${1}/{2}$)
&  0.2580  &  0.0754 &   0.0225   & 0.0068  &  0.0021  \\
% & 0.0006   & 0.0002& 0.0001  &  0.0000  &  0.0000  &  0.0000 &   0.0000  &  0.0000
\hline
CIPA
&0.2994 &    0.1010  &   0.0349   &  0.0122  &   0.0043  \\
%  & 0.0015   &  0.0005&  0.0002 &    0.0001   &  0.0000 &    0.0000   &  0.0000  &   0.0000\\
   \hline
 			\hline
     \multicolumn{6}{c}{\multirow{1}{*}{$M=2$}}\\ \hline
     \hline
 JPA(-${1}/{2}$, -${1}/{2}$)
 & 0.1860   & 0.0412  &  0.0098   & 0.0024 &   0.0006\\
 % &   0.0002  &  0.0000&     0.0000   & 0.0000  &  0.0000  &  0.0000   & 0.0000   & 0.0000\\
% 1.4142   & 0.3222 &   0.0838&    0.0227 &   0.0063   & 0.0017   & 0.0005&
% 0.0001   & 0.0000   & 0.0000&    0.0000  &  0.0000 &   0.0000\\
\hline
 JPA($\frac{1}{2}$, $\frac{1}{2}$)
& 0.1079  	&  0.0271   	& 0.0093   	& 0.0034   & 0.0012\\
% &   0.0005 &   0.0002& 0.0001&    0.0000 &   0.0000  &  0.0000  &  0.0000 &   0.0000\\
%1.4157  &   0.3589  &   0.1352 &   0.0524   &  0.0204 &    0.0079  &   0.0031&
% 0.0012   &  0.0005  &   0.0002  &   0.0000  &   0.0000  &   0.0000 \\
\hline
 JPA($\frac{1}{2}$, -$\frac{1}{2}$)
 & 0.0603  &  0.0056 &   0.0006  &  0.0001  &  0.0000 \\
 %&   0.0000   & 0.0000& 0.0000  &  0.0000  &  0.0000  &  0.0000  &  0.0000  &  0.0000
%1.4157 &    0.0912&     0.0076   &  0.0007 &    0.0001 &    0.0000   &  0.0000&
%0.0000   &  0.0000  &   0.0000  &   0.0000   &  0.0000   &  0.0000
\hline
JPA(0, -$\frac{1}{2}$)
& 0.0964 &   0.0123 &   0.0017  &  0.0003&    0.0000  \\
%&    0.0000   & 0.0000&  0.0000   & 0.0000  &  0.0000 &   0.0000   & 0.0000  &  0.0000
% 1.4152   & 0.1585 &   0.0208  &  0.0028 &   0.0004 &   0.0001  &  0.0000&
%     0.0000   & 0.0000  &  0.0000 &   0.0000   & 0.0000  &  0.0000
\hline
CIPA  &
0.1173   &   0.0193   &   0.0035 &     0.0007   &   0.0001  \\
%  &  0.0000   &   0.0000 &  0.0000    &  0.0000  &    0.0000   &   0.0000  &    0.0000 &     0.0000
%  1.4144  &  0.1780  &  0.0273  &  0.0045&    0.0008  &  0.0002  &  0.0000&
%  0.0000&    0.0000 &   0.0000  &  0.0000  &  0.0000  &  0.0000
     \hline\hline
     \multicolumn{6}{c}{\multirow{1}{*}{$M=3$}}\\ \hline
     \hline
 JPA(-${1}/{2}$, -${1}/{2}$)
 	&   0.0979   & 0.0113 &   0.0014 &  0.0002   & 0.0000 \\
 % &  0.0000   & 0.0000&   0.0000&    0.0000   & 0.0000 &   0.0000  &  0.0000   & 0.0000\\
%1.4140    &0.1652    &0.0218  &  0.0030  &  0.0004   & 0.0001  &  0.0000&
%0.0000&    0.0000   & 0.0000 &   0.0000  &  0.0000   & 0.0000\\
\hline
  JPA(${1}/{2}$, ${1}/{2}$)
	& 0.0581   &  0.0096  &   0.0022  &   0.0005  &   0.0001  \\
% &  0.0000 &    0.0000& 0.0000  &   0.0000  &   0.0000  &   0.0000  &   0.0000  &   0.0000\\
%    1.4136 &   0.2262  &  0.0553 &   0.0138  &  0.0035  &  0.0009  & 0.0002&
%    0.0001 &   0.0000  &  0.0000  &  0.0000  &  0.0000  &  0.0000  \\
\hline
  JPA(${1}/{2}$, -${1}/{2}$)
  &  0.0424  &   0.0021  &   0.0001  &   0.0000   &  0.0000 \\
  %&    0.0000  &   0.0000&  0.0000  &   0.0000  &  0.0000  &   0.0000  &   0.0000  &   0.0000
\hline
     JPA(0, -${1}/{2}$)
    & 0.0636  &   0.0046 &    0.0003  &   0.0000&     0.0000 \\
    % &   0.0000    & 0.0000 & 0.0000   & 0.0000  &  0.0000 &   0.0000   & 0.0000  &  0.0000
\hline
     CIPA
       & 0.0761   & 0.0067 &   0.0006 &   0.0001   & 0.0000 \\
       %&   0.0000   & 0.0000& 0.0000  &  0.0000  &  0.0000 &   0.0000  &  0.0000   & 0.0000
 \hline
\hline
\end{tabular}
\end{table}

\subsection{Denoising  stationary signals  on random geometric graphs}
\label{randomsignal.demo}

Let ${\bf L}^{\rm sym}$
be the normalized Laplacian on
the random geometric graph ${\mathcal G}_N$ with $N=256$.
In simulations of this subsection, we consider stationary signals ${\bf x}$  on the random geometric graph ${\mathcal G}_{256}$
with correlation  matrix  ${\mathbb E} {\bf x} {\bf x}^T={\bf I}+{\bf L}^{\rm sym}/2$,
 and noisy observations ${\bf y}={\bf x}+\pmb \epsilon$
being the  inputs ${\bf x}$ corrupted by some additive noises $\pmb \epsilon$ which
is independent of the input signal ${\bf x}$ and  whose entries  are i.i.d. random variables with normal distribution  ${\mathcal N}(0, \varepsilon)$ for some $\varepsilon>0$,
and  we select the  uniform probability measure $\bf P$ in the  stochastic
mean squared error \eqref{mse.objectivefunction}. In other words, we consider the Wiener filtering procedure \eqref{stochasticestimator}
in the scenario that
$${\bf H}={\bf I}, {\bf R}={\bf I}+{\bf L}^{\rm sym}/2, {\bf P}=N^{-1} {\bf I}\ {\rm and} \ {\bf G}=\varepsilon^2 {\bf I}.$$
For  input signals ${\bf x}$ in our simulations, one may verify
${\mathbb E}\|{\bf x}\|_2^2= {\rm tr} ({\mathbb E}({\bf  x}{\bf x}^T))=3N/2$,
${\mathbb E}\|{\pmb \epsilon}\|_2^2 %= {\rm tr} ({\mathbb E}({\pmb \eta}{\pmb \eta}^T))
=N \varepsilon^2$, and
\begin{eqnarray*} {\mathbb E}{\bf x}^T {\bf L}^{\rm sym}{\bf x} & \hskip-0.08in = & \hskip-0.08in  {\rm tr}\big( {\bf L}^{\rm sym} ({\bf I}+{\bf L}^{\rm sym}/2)\big)
%\nonumber\\
%& \hskip-0.08in = & \hskip-0.08in  N+\frac{1}{2} \Big(N+\sum_{i\in V} \sum_{(j,i)\in E}\frac{1}{d_id_j}\Big)
\in (3N/2, 2N].
 \end{eqnarray*}
 %where $d_i$ id the degree of the vertex $i\in V$.
 Based on the above observations, we use
 ${\bf K}=\varepsilon^2 {\bf L}^{\rm sym}/(4N)$
 as the regularization matrix
 to balance the fidelity and regularization terms
 in \eqref{mse.objectivefunction}.
Therefore
\begin{eqnarray*} {\bf x}_{\rm W0} & \hskip-0.08in := & \hskip-0.08in {\bf W}_{\rm mse}^0 {\bf y}={\bf R} \big({\bf R}+{\bf G}\big)^{-1} {\bf y}\nonumber\\
& \hskip-0.08in = & \hskip-0.08in
({\bf I}+{\bf L}^{\rm sym}/2)\big( (1+\varepsilon^2){\bf I}+ {\bf L}^{\rm sym}/2\big)^{-1}{\bf y}\end{eqnarray*}
 and
\begin{eqnarray*} {\bf x}_{\rm W} & \hskip-0.08in := & \hskip-0.08in {\bf W}_{\rm mse} {\bf y}  = ({\bf P}+{\bf K})^{-1}{\bf P} {\bf R} \big({\bf R}+{\bf G}\big)^{-1} {\bf y}\nonumber\\
& \hskip-0.08in = & \hskip-0.08in  ({\bf I}+\varepsilon^2 {\bf L}^{\rm sym}/4)^{-1} {\bf x}_{\rm W0}
% ({\bf I}+{\bf L}^{\rm sym}/2)\big( (1+\eta^2){\bf I}+ (\eta^4+\eta^2+1) {\bf L}^{\rm sym}/2+ \eta^2 (L^{\rm sym}/2)^2\big)^{-1}{\bf y}
\end{eqnarray*}  are
 signals   reconstructed from the  noisy observation ${\bf y}$
via  the  Wiener procedures \eqref{stochasticrecovery.eqa}  and  \eqref{wienerfiltermse.eq2} without/with regularization taken into account respectively.
%and the  Wiener procedure  with regularization.

\begin{figure}[t]
\centering
\includegraphics[width=43mm, height=30mm]{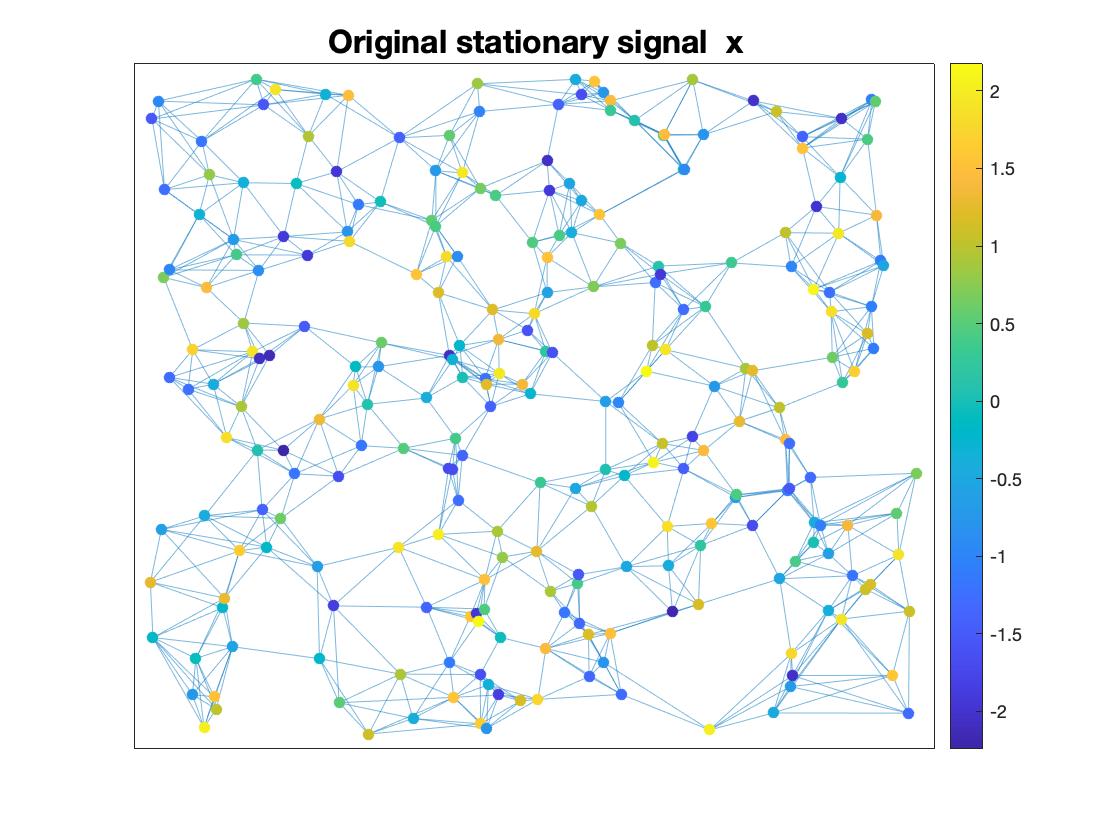}
\includegraphics[width=43mm, height=30mm]
{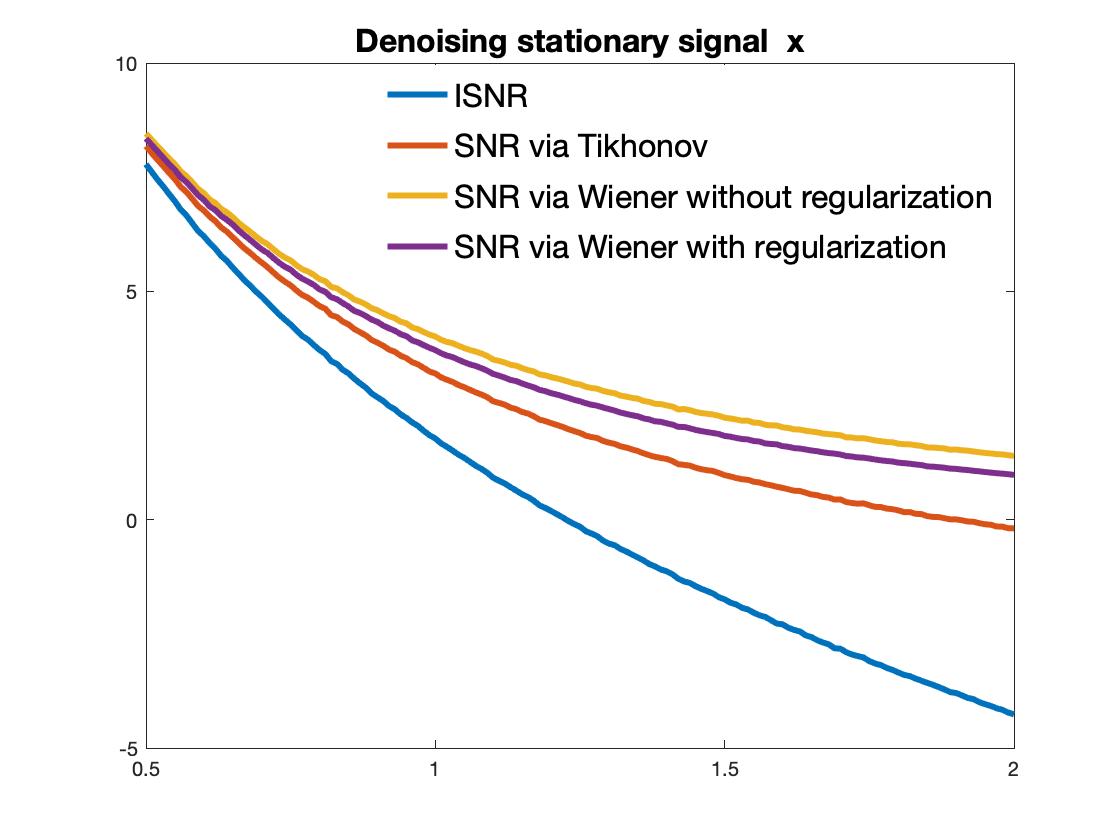}
\\
\includegraphics[width=42mm, height=30mm]{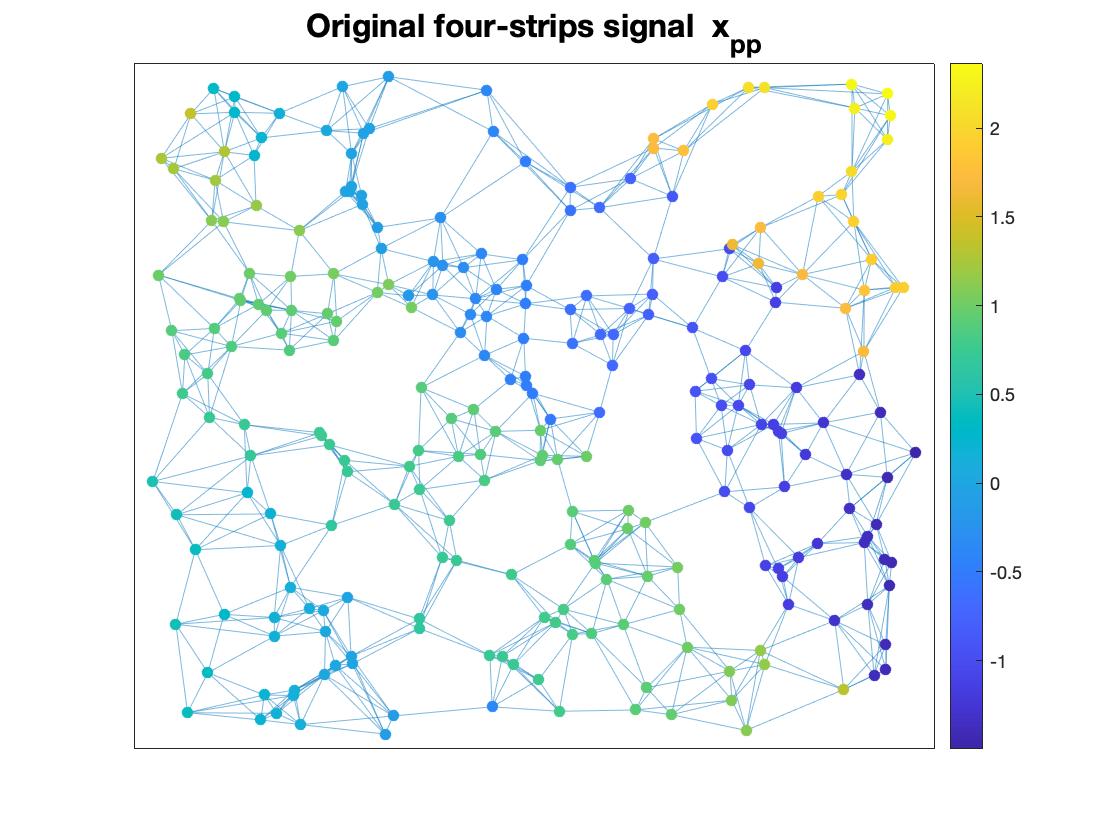}
\includegraphics[width=43mm, height=30mm] {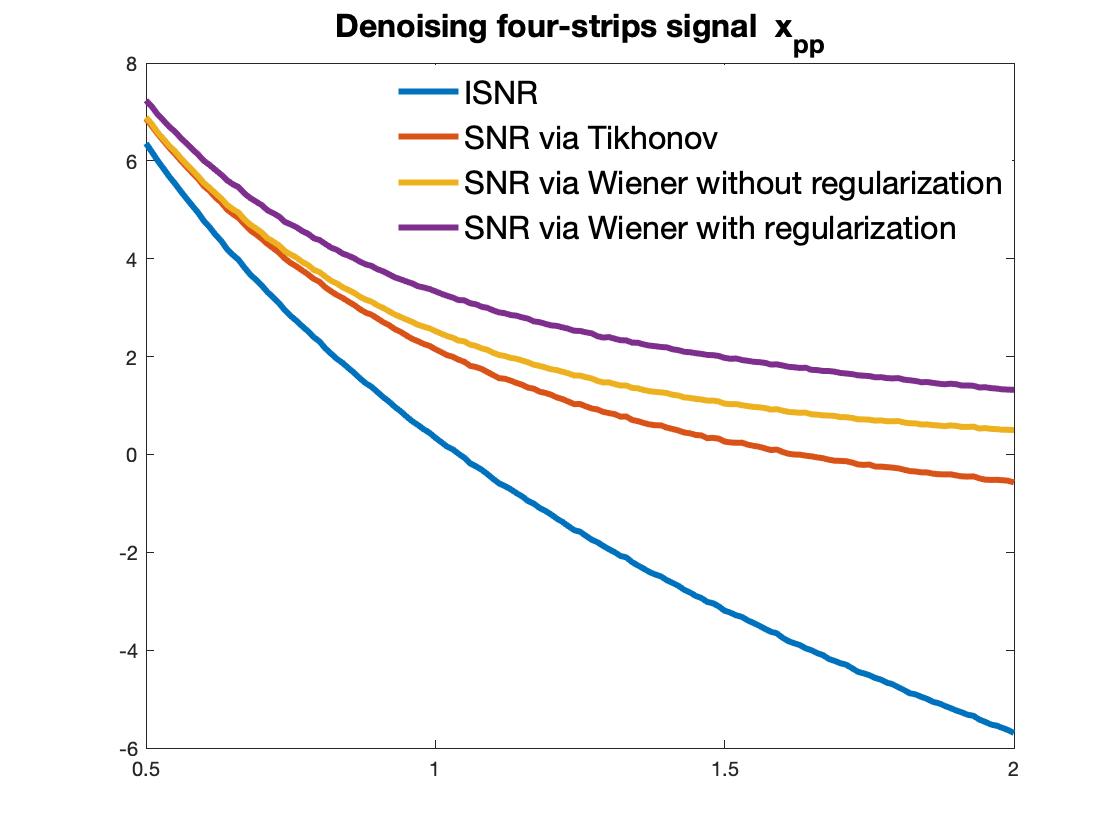}
\caption{ Plotted  are the stationary signal ${\bf x}$ with correlation matrix  ${\bf I}+{\bf L}^{\rm sym}/2$ (top left),
the four-strip signal  ${\bf x}_{\rm pp}$  in \cite{jiang19} (bottom left),
and  the averages of the
input signal-to-noise ratio
${\rm ISNR}$ and
output signal-to-noise ratio ${\rm SNR}$
of denoising  stationary signals ${\bf x}$ (top right) and  the four-strip signal ${\bf x}_{\rm pp}$   (bottom right)
via the  Wiener procedures  without/with regularization and   Tikhonov regularization approach
 over 1000 trials for different noise levels $0.5 \le \varepsilon\le  2$.
 }
\label{denoise_random.fig}
\end{figure}

Define  the input signal-to-noise ratio
(ISNR) and
the output signal-to-noise ratio (SNR) by
	$$
{\rm ISNR}= -20 \log_{10} \frac{\|{\pmb \epsilon}\|_2}{\|{\bf x}\|_2}\ {\rm and} \
{\rm SNR}=-20 \log_{10} \frac{\|\widehat{\bf x}-{\bf x}\|_2}{\|{\bf x}\|_2} $$
respectively, 	where $\widehat{\bf x}$ are either the reconstructed signal ${\bf x}_{\rm W0}$
via  the  Wiener procedure \eqref{stochasticrecovery.eqa} without regularization, or
the reconstructed signal ${\bf x}_{\rm W}$ via
 the  Wiener procedure \eqref{wienerfiltermse.eq2} with regularization, or the reconstructed signal
\begin{eqnarray}\label{tik.def}
{\bf x}_{\rm Tik} & \hskip-0.08in = & \hskip-0.08in ({\bf P}+{\bf K})^{-1}{\bf P} {\bf y}= ({\bf I}+\varepsilon^2 {\bf L}^{\rm sym}/2)^{-1} {\bf y}\nonumber\\
&\hskip-0.08in = & \hskip-0.08in \arg\min_{\bf x} \ ({\bf x}-{\bf y})^T {\bf P} ({\bf x}-{\bf y})+  {\bf x}^T {\bf K} {\bf x}
\end{eqnarray}
via the  Tikhonov regularization approach.
 It is observed from  Figure \ref{denoise_random.fig} that the Wiener procedure without  regularization has the best performance on
 denoising stationary signals.

 Graph signals  ${\bf x}$ in many applications exhibit some smoothness,  which is widely measured  by
the ratio ${\bf x}^T {\bf L}^{\rm  sym} {\bf x}/ \|{\bf x}\|_2^2$. % \cite{ ekambaram13, Grassi18, horn1990matrix}.
Observe that stationary signals ${\bf x}$ in the above simulations does not have good  regularity as
${\mathbb E}{\bf x}^T {\bf L}^{\rm  sym} {\bf x}/ {\mathbb E}\|{\bf x}\|_2^2\in [1, 4/3]$. We believe  that it  could be the reason that
Wiener procedure with regularization has slightly poor performance on denoising than   the Wiener procedure without regularization  does.

Let   ${\bf x}_{\rm pp}$ be  the four-strip signal on the random geometric graph
that impose the polynomial $0.5-2 c_x$ on the first and third diagonal strips and $0.5+c_x^2+c_y^2$ on the second
and fourth strips respectively, where $(c_x,c_y)$ are the coordinates of vertices \cite[Fig. 2]{jiang19}.
We do simulations %\st{tested the performance of the
%Wiener procedure with/without regularization}
on denoising the four-strip signal ${\bf x}_{\rm pp}$,
%on the random geometric graph ${\mathcal G}_{256}$,
i.e., we apply the same
Tikhonov regularization and Wiener procedures with/without regularization except  that stationary signals
${\bf x}$ is replaced by ${\bf x}_{\rm pp}$, see Figure \ref{denoise_random.fig}. %{denoise_piecewise.fig}.
This  indicates that
 Wiener procedure with regularization may  have the best performance on denoising signals with certain regularity.

\subsection{Denoising wide-band stationary signals  on random geometric graphs}
\label{denoisingwideband.demo}
\begin{figure}[t] %[h]
\centering
\includegraphics[width=43mm, height=30mm]
{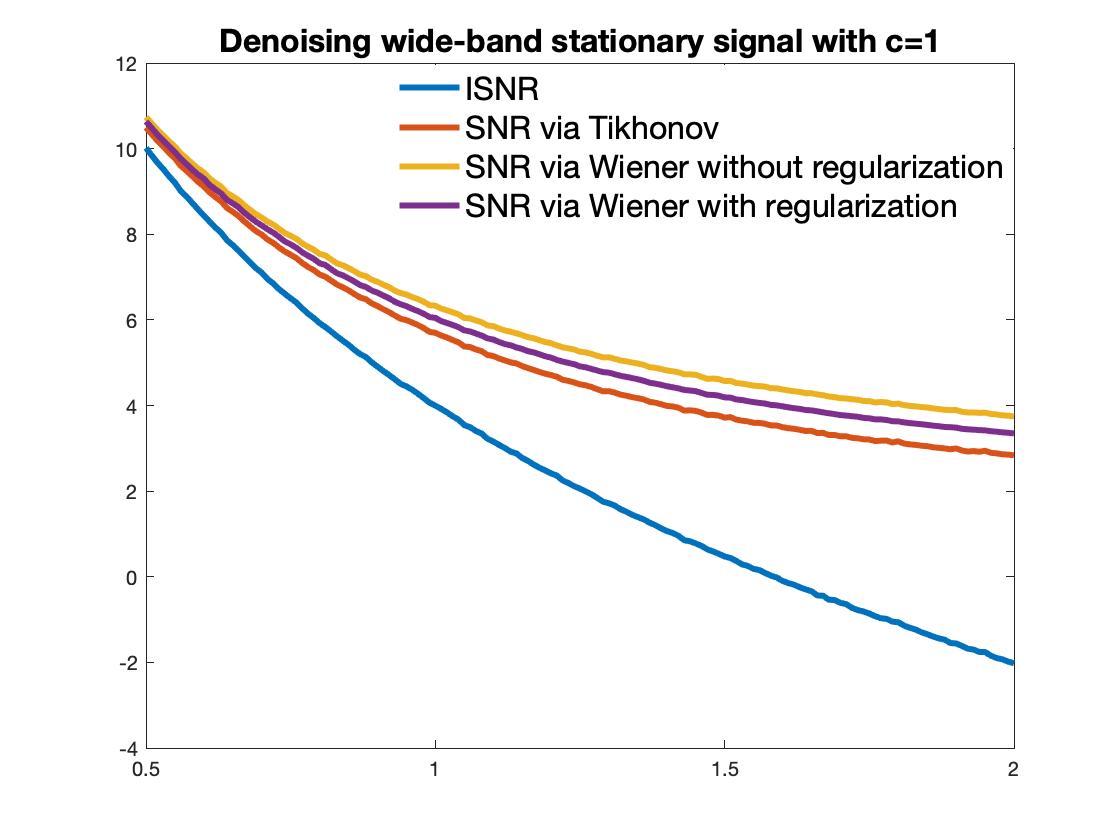}
\includegraphics[width=43mm, height=30mm]
{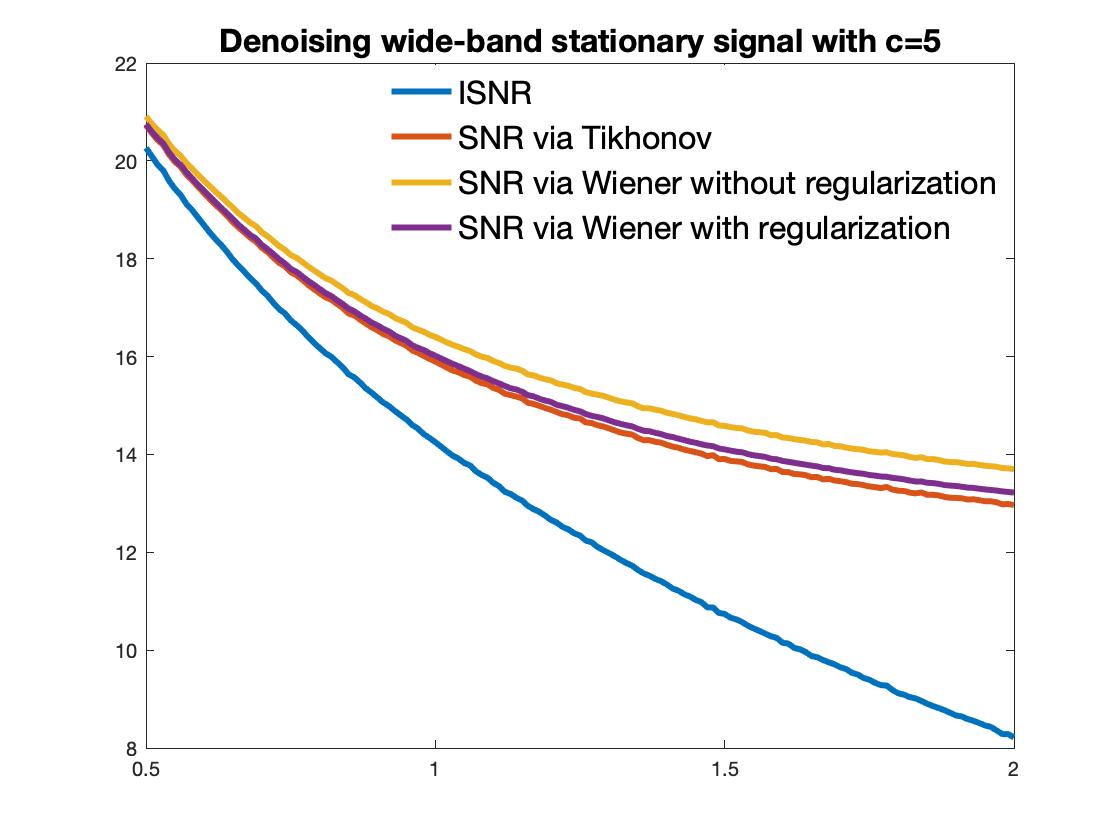}
\caption{Plotted are the averages of the input signal-to-noise ratio ${\rm ISNR}$ and  output signal-to-noise ratio ${\rm SNR}$  obtained by the Wiener procedures without/with regularization  and Tikhonov regularization approach over 1000 trials for different noise levels $0.5\le \varepsilon\le  2$, in which the original signal is wide-band stationary  with $c=1$ (left) and $c=5$ (right) on the random geometric graph ${\cal G}_{256}$.
}
\label{stationary.fig}
\end{figure}

In this subsection, we consider denoising  wide-band stationary  signals ${\bf x}$ in \eqref{wiener2.eq4} on a random geometric graph  ${\cal G}_{256}$ with
$${\mathbb E}{\bf x}= c{\bf 1}\ \ {\rm and}\ \
{\mathbb E} ({\bf x}-{\mathbb E}{\bf x}) ({\bf x}-{\mathbb E}{\bf x}) ^T={\bf I}+{\bf L}^{\rm sym}/2,$$
where  $c\ne 0$ is not necessarily to be given in advance.  The observations ${\bf y}={\bf x}+\pmb \epsilon$
are the  inputs ${\bf x}$ corrupted by some additive noises $\pmb \epsilon$ which
is independent of the input signal ${\bf x}$ and  whose covariance matrix is ${\bf G}=\varepsilon^2{\bf L}^{\rm sym}$ for  some $\varepsilon>0$,
%correlation matrix ${\mathbb E}({\pmb \epsilon}{\pmb \epsilon}^T)= \varepsilon {\bf I}$ for some $\epsilon>0$,
% it has mean zero $0$, and covariance
%matrix  ${\bf G}=\eta^2$,
and  we select the  uniform probability measure $\bf P$ in the  stochastic
mean squared error. In other words, we consider the Wiener filtering procedure \eqref{stochasticestimator}
in the scenario that
$${\bf H}={\bf I}, \widetilde {\bf R}={\bf I}+{\bf L}^{\rm sym}/2, {\bf P}=N^{-1} {\bf I} \ {\rm and} \ {\bf G}=\varepsilon^2 {\bf L}^{\rm sym}.$$
Similar to the simulations in Section \ref{randomsignal.demo}, we test the performance of the
Wiener procedures with/without regularization and Tikhonov regularization on denoising  wide-band stationary signals.
From the simulation results presented in  Figure \ref{stationary.fig}, we see that the Wiener procedure with regularization has slightly poor performance on denoising than  the Wiener procedure without regularization  does, but they  both perform better than Tikhonov regularization approach does.

\appendices
\setcounter{equation}{0}
\setcounter{thm}{0}
\setcounter{section}{0}
\setcounter{subsection}{0}
\renewcommand{\thesection}{%Appendix
\Alph{section}}
\renewcommand{\thesubsection}{A.\arabic{subsection}}
\renewcommand{\theequation}{\Alph{section}.\arabic{equation}}

\section{Proof of Theorem \ref{wienerfiltermse.thm}}
\label{wienerfiltermsd.prof}

%Let ${\bf  e}_i , i \in  V$, be the delta signals taken value one at vertex
%$i$ and zero at all other vertices.
  By \eqref{wiener2.eq1}, \eqref{wiener2.eq2} and  \eqref{wiener2.eq3}, we have
  \begin{equation} \label{wienerfiltermse.thmpf.eq0} {\mathbb E} {\bf y}{\bf y}^T= {\bf H} {\bf R} {\bf H}^T+{\bf G}
 \ \  {\rm and} \  \ {\mathbb E} {\bf y}{\bf x}^T={\bf H}{\bf R}.
   \end{equation}
   By \eqref{wiener2.eq1}, \eqref{mse.objectivefunction} and
  \eqref {wienerfiltermse.thmpf.eq0},
we obtain
\begin{eqnarray} \label{wienerfiltermse.thmpf.eq1}  & \hskip-0.08in & \hskip-0.08in
F_{{\rm mse}, P, {\bf K}}({\bf W})\nonumber\\
% & \hskip-0.08in = & \hskip-0.08in
%{\mathbb E} ( ({\bf W}{\bf y}-{\bf x})^T{\bf P} ({\bf W}{\bf y}-{\bf x})+ {\bf y}^T{\bf W}^T {\bf K}{\bf W}{\bf y})
% \nonumber \\
 & \hskip-0.08in = & \hskip-0.08in
 {\rm tr}\left (  {\bf P} {\mathbb E} ( ({\bf W}{\bf y}-{\bf x})({\bf W}{\bf y}-{\bf x})^T \right)
 +{\rm tr} {\bf W}^T {\bf K}{\bf W}{\mathbb E}({\bf y} {\bf y}^T) \nonumber\\
 & \hskip-0.08in = & \hskip-0.08in
 {\rm tr} \big({\bf W}^T ({\bf P}+{\bf K}) {\bf W} ({\bf H} {\bf R} {\bf H}^T+{\bf G})\big) +{\rm tr} ({\bf P} {\bf R})
 \nonumber\\
 & & \hskip-0.08in -  {\rm tr} ( {\bf H}{\bf R}{\bf P}{\bf W})   - {\rm tr} ({\bf W}^T {\bf P}{\bf R} {\bf H}^T ).
% sum_{i\in V} p(i) \Big [ {\bf e}_i^T ({\bf W}{\bf H}-{\bf I})
%\big( {\mathbb E} {\bf x}{\bf x}^T\big) ({\bf W}{\bf H}-{\bf I})^T {\bf e}_i
%\nonumber\\
%& \hskip-0.08in & \hskip-0.08in
%  + 2   {\bf e}_i^T ({\bf W}{\bf H}-{\bf I})
%\big( {\mathbb E} {\bf x}{\pmb \epsilon}^T\big) {\bf W}^T {\bf e}_i+ {\bf e}_i^T {\bf W}
%\big( {\mathbb E} {\pmb \epsilon}{\pmb \epsilon}^T\big) {\bf W}^T {\bf e}_i\Big]  \nonumber
%\\
%& \hskip-0.08in= & \hskip-0.08in
%{\rm tr}
%\left( {\bf P} ({\bf W}{\bf H}-{\bf I}){\bf R}({\bf W}{\bf H}-{\bf I})^T +{\bf P}{\bf W} {\bf G} {\bf W}^T \right).
%\nonumber\\
%& \hskip-0.08in= & \hskip-0.08in
%{\rm tr} \left ({\bf P} {\bf W} ({\bf H}{\bf R} {\bf H}^T +{\bf G}) {\bf W}^T-  {\bf P}{\bf W} {\bf H}{\bf R}- {\bf P}{\bf R} {\bf H}^T {\bf W}^T+ {\bf P} {\bf R}\right).
\end{eqnarray}
Substituting ${\bf W}$ in \eqref{wienerfiltermse.thmpf.eq1}  by ${\bf W}_{\rm mse}$
%, we conclude that
%$$F_{{\rm mes}, P}({\bf W}_{\rm mse})=
%{\rm tr} \left({\bf P}{\bf R}-{\bf P} {\bf R} {\bf H}^T ({\bf H}{\bf R} {\bf H}^T +{\bf G})^{-1} {\bf H}{\bf R}\right)$$
%which
proves \eqref{wienerfiltermse.eq1}.

By \eqref{wienerfiltermse.eq2} and \eqref{wienerfiltermse.thmpf.eq1}, we obtain
\begin{eqnarray} \label{wienerfiltermse.thmpf.eq2}
& \hskip-0.08in  & %\hskip-0.08in
F_{{\rm mse}, P, {\bf K}}({\bf W})    \nonumber\\
& \hskip-0.08in  = & \hskip-0.08in  F_{{\rm mse}, P, {\bf K}}({\bf W}_{\rm mse}) +{\rm tr} \big({\bf V}^T ({\bf P}+{\bf K}) {\bf V} ({\bf H} {\bf R} {\bf H}^T+{\bf G})\big)
  \nonumber\\
& & +
{\rm tr} \big({\bf V}^T ({\bf P}+{\bf K}) {\bf W}_{\rm mse} ({\bf H} {\bf R} {\bf H}^T+{\bf G})-{\bf V}^T {\bf P}{\bf R} {\bf H}^T\big) \nonumber\\
 & &
 + {\rm tr} \big({\bf W}_{\rm mse}^T ({\bf P}+{\bf K}) {\bf V} ({\bf H} {\bf R} {\bf H}^T+{\bf G})-  {\bf H}{\bf R}{\bf P}{\bf V}\big)
  \nonumber\\
 % & &
%+ {\rm tr} \left( {\bf P} {\bf V}  (({\bf H}{\bf R} {\bf H}^T +{\bf G}) {\bf W}_{\rm mse}^T -{\bf H}{\bf R})\right)\nonumber\\
%& & +
%{\rm tr} \left( {\bf P}  ( {\bf W}_{\rm mse}({\bf H} {\bf R} {\bf H}^T +{\bf G})  -{\bf R}{\bf H}^T)  {\bf V}^T\right) \nonumber\\
& \hskip-0.08in = & \hskip-0.08in   F_{{\rm mse}, P, {\bf K}}({\bf W}_{\rm mse})  +{\rm tr} \big(({\bf H} {\bf R} {\bf H}^T+{\bf G})^{1/2}\nonumber\\
 & & \qquad\qquad  \times {\bf V}^T ({\bf P}+{\bf K}) {\bf V} ({\bf H} {\bf R} {\bf H}^T+{\bf G})^{1/2}\big)
\nonumber\\
& \hskip-0.08in \ge  & \hskip-0.08in  F_{{\rm mse}, P, {\bf K}}({\bf W}_{\rm mse}),
\end{eqnarray}
where  ${\bf V}={\bf W}-{\bf W}_{\rm mse}$, the first and second equality follows from \eqref{wienerfiltermse.thmpf.eq1} and \eqref{wienerfiltermse.eq2} respectively,
and the  inequality holds as
 $({\bf H} {\bf R} {\bf H}^T+{\bf G})^{1/2}{\bf V}^T ({\bf P}+{\bf K}) {\bf V} ({\bf H} {\bf R} {\bf H}^T+{\bf G})^{1/2}$ are
  positive semidefinite for all matrices ${\bf V}$.
This proves that ${\bf W}_{\rm mse}$ is a minimizer to the minimization problem $\min_{\bf W} F_{{\rm mse}, P, {\bf K}}({\bf W})$.

The conclusion that  ${\bf W}_{\rm mse}$ is a unique minimizer to the minimization problem $\min_{\bf W} F_{{\rm mse}, P, {\bf K}}({\bf W})$
follows from \eqref{wienerfiltermse.thmpf.eq2} and the assumptions that
${\bf P}+{\bf K}$ and ${\bf H} {\bf R} {\bf H}^T+{\bf G}$ are strictly positive definite.

\section{Proof of Theorem \ref{wienerfilterworsecase.thm}}
\label{wienerfilterworsecase.thm.pfappendix}

%\begin{proof}[Proof of Theorem \ref{wienerfilterworsecase.thm}]
Define the worst-case mean squared error
of a reconstruction vector ${\bf w}$ with  respect to a given unit vector ${\bf u}$ by
\begin{equation} \label{wienerfilterworsecase.thm.pfeq1}
f_{{\rm wmse}, {\bf u}}({\bf w})=\max_{\|{\bf x}\|_2\le \delta_0}
{\mathbb E}  | {\bf w}^T {\bf y}-{\bf u}^T {\bf x}|^2\end{equation}
and set
 \begin{equation} \label{wienerfilterworsecase.thm.pfeq2}
{\bf w}_{{\rm wmse}, {\bf u}}=  {\bf W}_{{\rm wmse}}^T % \delta_0^2 \big(\delta_0^2 {\bf H}{\bf H}^T+{\bf G})^{-1}{\bf H}
{\bf u}.
\end{equation}
By direct computation, we have
\begin{equation}  \label{wienerfilterworsecase.thm.pfeq1+}
F_{{\rm wmse}, P}({\bf W})=\sum_{i\in V} p(i) f_{{\rm wmse}, {\bf e}_i}( {\bf W}^T {\bf e}_i), \end{equation}
where
${\bf e}_i, i\in V$, are   delta signals taking value one at vertex $i$  and zero at all other vertices.
Then it suffices to  show that
${\bf w}_{{\rm wmse}, {\bf u}}$
is the optimal reconstructing vector with respect to the measurement $f_{{\rm wmse}, {\bf u}}({\bf w})$, i.e.,  % by solving the minimization problem
\begin{equation}
\label{wienerfilterworsecase.thm.pfeq3}
{\bf w}_{{\rm wmse}, {\bf u}}=\arg\min_{\bf w} f_{{\rm wmse}, {\bf u}}({\bf w}).
\end{equation}

By \eqref{wiener1.eqc}, \eqref{wiener1.eqd} and the assumption $\|{\bf u}\|_2=1$, we have
\begin{eqnarray*} \label{wienerfilterworsecase.thm.pfeq4}
\hskip-0.18in f_{{\rm wmse}, {\bf u}}({\bf w})
& \hskip-0.08in = &  \hskip-0.08in   \max_{\|{\bf x}\|_2\le \delta_0}
{\mathbb E}  | ({\bf w}^T {\bf H}-{\bf u}^T) {\bf x} + {\bf w}^T  {\pmb \epsilon}|^2\nonumber \\
\hskip-0.18in &\hskip-0.08in  = & \hskip-0.08in  \max_{\|{\bf x}\|_2\le \delta_0}\big| ({\bf w}^T {\bf H}-{\bf u}^T) {\bf x} \big|^2+
{\mathbb E} | {\bf w}^T  {\pmb \epsilon}|^2\nonumber\\
\hskip-0.18in & \hskip-0.08in  = & \hskip-0.08in  \delta_0^2 ({\bf w}^T {\bf H}-{\bf u}^T)
({\bf H}^T {\bf w}-{\bf u})
 +
{\bf w}^T {\bf G}{\bf w} \nonumber\\
\hskip-0.18in & \hskip-0.08in  = & \hskip-0.08in  {\bf w}^T  \big(\delta_0^2 {\bf H}{\bf H}^T+{\bf G}) {\bf w}-2 \delta_0^2 {\bf w}^T {\bf H} {\bf u}+\delta_0^2.
\end{eqnarray*}
Therefore
\begin{eqnarray} \label{wienerfilterworsecase.thm.pfeq5}
\hskip-0.18in  f_{{\rm wmse}, {\bf u}}({\bf w})
  & \hskip-0.08in = &\hskip-0.08in  f_{{\rm wmse}, {\bf u}}({\bf w}_{{\rm wmse}, {\bf u}})+ {\bf v}^T \big(\delta_0^2 {\bf H}{\bf H}^T+{\bf G}) {\bf v}\nonumber\\
\hskip-0.18in  & \hskip-0.08in  & + 2 {\bf v}^T \Big( \big(\delta_0^2 {\bf H}{\bf H}^T+{\bf G}\big)  {\bf w}_{{\rm wmse}, {\bf u}}-\delta_0^2{\bf H} {\bf u}\Big)\nonumber\\
\hskip-0.18in  & \hskip-0.08in = &\hskip-0.08in f_{{\rm wmse}, {\bf u}}({\bf w}_{{\rm wmse}, {\bf u}})+  {\bf v}^T \big(\delta_0^2 {\bf H}{\bf H}^T+{\bf G}) {\bf v}\nonumber\\
\hskip-0.18in  & \hskip-0.08in \ge &\hskip-0.08in   f_{{\rm wmse}, {\bf u}}({\bf w}_{{\rm wmse}, {\bf u}}),
\end{eqnarray}
where ${\bf v}={\bf w}-{\bf w}_{{\rm wmse}, {\bf u}}$ and the last inequality holds as
$\delta_0^2 {\bf H}{\bf H}^T+{\bf G}$ is strictly positive definite.
This proves \eqref{wienerfilterworsecase.thm.pfeq3} and hence
that
${\bf W}_{\rm wmse}$ is a minimizer of the minimization problem
\eqref{wcms.minimization}, i.e., the inequality in \eqref{wienerfilterworsecase.eq1} holds.

By \eqref{wienerfilterworsecase.thm.pfeq2}, \eqref{wienerfilterworsecase.thm.pfeq1+}  and \eqref{wienerfilterworsecase.thm.pfeq3}, we have
%the following estimate to
% the worst-case mean squared error  for the worst-case Wiener filter ${\bf W}_{\rm wmse}$,
\begin{eqnarray*}%\label{wienerfilterworsecase.error}
 &\hskip-0.08in  &  \hskip-0.08in F_{{\rm wmse}, P}({\bf W}_{\rm wmse})  =  \sum_{i\in V} p(i) f_i({\bf w}_{{\rm wmse}, {\bf e}_i})\nonumber\\
&  \hskip-0.08in = & \hskip-0.08in \sum_{i\in V} p(i) \big(-\delta_0^4 {\bf e}_i^T {\bf H}^T (\delta_0^2 {\bf H} {\bf H}^T +{\bf G})^{-1} {\bf H} {\bf e}_i+\delta_0^2\big)
\nonumber\\
&  \hskip-0.08in = & \hskip-0.08in  \delta_0^2- \delta_0^4 {\rm tr} ( {\bf P} {\bf H}^T (\delta_0^2 {\bf H} {\bf H}^T +{\bf G})^{-1} {\bf H}\big)\nonumber\\
&  \hskip-0.08in = & \hskip-0.08in  \delta_0^2- \delta_0^4 {\rm tr} ( (\delta_0^2 {\bf H} {\bf H}^T +{\bf G})^{-1} {\bf H} {\bf P} {\bf H}^T \big).
\end{eqnarray*}
This proves the equality in \eqref{wienerfilterworsecase.eq1} and hence completes the proof of
the conclusion \eqref{wienerfilterworsecase.eq1}.

The uniqueness of the minimization problem
\eqref{wcms.minimization} follows from
 \eqref{wienerfilterworsecase.thm.pfeq1+} and \eqref{wienerfilterworsecase.thm.pfeq5},  and the strictly positive definiteness  of the matrices $\bf P$ and
 $\delta_0^2 {\bf H}{\bf H}^T+{\bf G}$.

\medskip

{\bf Acknowledgement}\  The authors would like to thank Professors  Xin Li, Zuhair Nashed, Paul Nevai and Yuan Xu, and Dr. Nazar Emirov for their help during the
preparation of this manuscript.

\bibliographystyle{ieeetr}

%\bibliography{ref}
\end{document}